\documentclass[11pt]{article} 
\usepackage[utf8]{inputenc}
\usepackage{multirow}
\usepackage{amsfonts}
\usepackage{amsmath}
\usepackage{amssymb}
\usepackage{amsthm}
\usepackage{caption}
\usepackage[dvipsnames]{xcolor}
    \definecolor{darkblue}{rgb}{0,0.5,0}
    \definecolor{darkblue}{rgb}{0,0,0.6}
    \definecolor{purple}{rgb}{0.4,.2,0.7}
\usepackage[margin = 2.5cm]{geometry}
\usepackage{graphicx}
\usepackage[hyperfootnotes = false, colorlinks = true, linkcolor = darkblue, citecolor = purple]{hyperref}
\usepackage{subcaption}
\usepackage{ytableau}

\usepackage{csquotes}

\usepackage{tikz}
\usetikzlibrary{decorations.pathmorphing}
\usetikzlibrary{decorations.markings}
\definecolor{mathblue}{RGB}{180,44,37}
\definecolor{mathblue}{RGB}{39,94,190}
\tikzset{>=latex} % for LaTeX arrow head
\tikzset{ photon/.style={decorate, decoration={snake}, draw=black}}

\pdfoutput = 1
    
    \usepackage[utf8]{inputenc}
    \usepackage{color,graphicx}
    \usepackage{epsfig}
    \usepackage{amsmath}
    \usepackage{verbatim}
    \usepackage{amssymb}
    \usepackage{mathabx}
    \usepackage[mathcal]{eucal}
    \usepackage{stmaryrd}
    \usepackage{empheq}
    \usepackage{esint}
    \usepackage{physics}
    \usepackage{amsfonts}
    \usepackage{bbm}
    \usepackage{array}
    \usepackage{setspace}
    \usepackage{braket}
    \usepackage{float}
    \usepackage{url}
    \usepackage{mathtools}
    \usepackage{tensor}
    \usepackage{bbold}
    \usepackage{slashed}
    \usepackage{tikz}
    \usepackage{graphicx}
    \usepackage{epstopdf}
    \usepackage{subcaption}
    \usepackage[labelfont=bf,font={sf}]{caption}
    \usepackage{pgfplots}
    \usepackage{mathrsfs}
    \usepackage{fancybox}
    \usepackage{eurosym}
    \usepackage{tcolorbox}
    \usepackage{tensor}
    \usepackage[normalem]{ulem}
    \allowdisplaybreaks
    \usepackage{changepage}

    \usepackage[ragged]{footmisc}
    \newcommand{\eqcref}[1]{Eq.~\eqref{#1}}

    \def\ben{\begin{equation}}
    \def\een{\end{equation}}

         \let\r=v

    \def\be{\begin{equation}}
    \def\ee{\end{equation}}
    \def\ba{\begin{array}}
    \def\ea{\end{array}}

    \def\dalemb#1#2{{\vbox{\hrule height .#2pt
           \hbox{\vrule width.#2pt height#1pt \kern#1pt
                   \vrule width.#2pt}
           \hrule height.#2pt}}}
    
    \newcommand{\bea}{\begin{eqnarray}}
    \newcommand{\eea}{\end{eqnarray}}

    \let\tilde=\widetilde

    \numberwithin{equation}{section}
    
    \usetikzlibrary{decorations.pathmorphing} % For zigzag

\usepackage{comment}

\usepackage{physics}

\begin{document}

\thispagestyle{empty}
\begin{center}
    ~\vspace{5mm}

  \vskip 2cm 
  
   {\LARGE \bf 
       Stringy algebras, stretched horizons, and quantum-connected wormholes
   }

   \vspace{0.5in}
     
   {\bf Aidan Herderschee and Jonah Kudler-Flam
   }

    \vspace{0.5in}

School of Natural Sciences, Institute for Advanced Study, Princeton, NJ, USA               
    \vspace{0.5in}

    \vspace{0.5in}
    
%     {\tt   malda@ias.edu}

\end{center}

\vspace{0.5in}

    \begin{abstract}
     While the supergravity limit of AdS/CFT has been extensively explored, the regime in which stringy dynamics dominate, characterized by the emergence of an infinite tower of higher-spin massive modes, is far less understood. In this work, we leverage techniques from algebraic quantum field theory to investigate the extent to which hallmark features of bulk gravity survive at finite string tension and the emergence of intrinsically stringy phenomena. Working in the $g_s\rightarrow 0$ limit, we model excited string modes as free particles and demonstrate that the resulting Hagedorn spectrum leads to the breakdown of the split property, a strengthening of the locality principle, for regions that are within a string length of each other. We propose that this leads to a precise algebraic definition of stretched horizons and stretched quantum extremal surfaces. When stretched horizons exist, there is an associated nontrivial horizon $\star$-algebra.
      Furthermore, we investigate the algebraic ER=EPR proposal, describing the conditions for the emergence of type III von Neumann factors of all subtypes, which provides an intriguing characterization of how such regions can have a quantum Einstein-Rosen bridge even if they are geometrically disjoint.

    \end{abstract}

    \pagebreak
    \setcounter{page}{1}
    \setcounter{tocdepth}{2}
    % \tableofcontents
    
    %%%%%%%%%%%
    % SECTION %
    %%%%%%%%%%%

\tableofcontents

\section{Introduction}

The AdS/CFT correspondence posits that strongly coupled gauge theories encode the full dynamics of a higher-dimensional gravitational theory in asymptotically anti–de Sitter (AdS) space \cite{Maldacena:1997re,Witten:1998qj,Gubser:1998bc}. Hallmark gravitational phenomena acquire sharp field-theoretic counterparts. For instance, the Hawking-Page phase transition where black holes dominate the canonical ensemble \cite{Hawking:1982dh} corresponds to the deconfinement transition \cite{Witten:1998zw}, areas of codimension-two regions of spacetime correspond to quantum entanglement \cite{Ryu:2006bv,Ryu:2006ef}, and the scattering of shock waves near black hole horizons corresponds to quantum chaos \cite{Shenker:2013pqa,Shenker:2014cwa,Maldacena:2015waa}. Spurred by the discovery of Liu and Leutheusser that novel operator algebras emerge in the large-$N$ limit of holographic conformal field theories \cite{Leutheusser:2021qhd,Leutheusser:2021frk}, there has been a resurgence of focus on von Neumann algebras to sharpen our understanding of holography. This has led to progress in understanding the interior of black holes \cite{Leutheusser:2021qhd,Leutheusser:2021frk}, generalized entropy in both AdS/CFT \cite{Witten:2021unn, Chandrasekaran:2022eqq} and cosmological spacetimes \cite{Chandrasekaran:2022cip,Jensen:2023yxy, Witten:2023xze, Kudler-Flam:2023qfl,Chen:2024rpx,Kudler-Flam:2024psh,Blommaert:2025bgd}, the generalized second law \cite{Chandrasekaran:2022eqq, Faulkner:2024gst, Kirklin:2024gyl}, quantum chaos and information loss \cite{Furuya:2023fei, Gesteau:2023rrx, Ouseph:2023juq, Kolchmeyer:2024fly, Penington:2025hrc}, the ER=EPR proposal \cite{Engelhardt:2023xer}, and the emergence of horizons \cite{Gesteau:2024rpt,Kolchmeyer:2024fly}.

The vast majority of progress has been made in the semi-classical regime, where the string tension is parametrically large and the string coupling is parametrically small. This is the limit in which quantum field theory in curved spacetime is a good approximation. Our goal in this paper is to understand the regime where the string tension is finite, allowing the string to become a finite-size object. We therefore must reckon with the inherent non-locality of the theory. For the majority of our work, we will still consider the zero string coupling limit. In section \ref{sec:discus}, we will back away from this limit by considering some perturbative corrections.
These questions are not of merely formal interest. They are directly relevant to the regime that prospective future experiments are most likely to probe, namely the BFSS model \cite{Banks:1996vh} in the regime dual to type IIA string theory at finite string coupling \cite{Itzhaki:1998dd,Maldacena:2023acv}.

We are led to apply the theoretical framework of string field theory \cite{Erbin:2021smf,Sen:2024nfd}
to investigate models with finite string tension.
In the limit of vanishing string coupling, so interactions between strings can be neglected, the string field theory reduces to an infinite collection of free particles with the same quantum numbers as the excited states of the string.\footnote{This approximation is well supported by the fact that it reproduces the correct free energy \cite{Polchinski:1985zf,Maldacena:2000kv}. One might still worry that this approximation is inapplicable for the phenomena we are studying because it appears to give the wrong value for the cosmological constant \cite{Polchinski:1985zf} and thus potentially other phenomena. We regard the mismatch as misleading, as the cosmological constant should be treated as part of the definition of the local algebra rather than as a prediction. We are mainly concerned with the non-locality of the theory, which is present from the infinite tower of states.}
This limit introduces some essential aspects of stringy non-locality while still being solvable.

The crucial feature of this limit is that the string's excited modes exhibit a Hagedorn spectrum, i.e.~an exponential dependence on the energy in the density of states, $\rho(E) \sim e^{\beta_H E}$ \cite{DelGiudice:1971yjh}. Theories with Hagedorn spectra have a maximal temperature $\beta_H$, above which thermodynamics is ill-defined because the partition function diverges, usually signaling a phase transition.\footnote{In general, one expects that a phase transition to occur slightly before or after the Hagedorn temperature in string theory \cite{Chen:2021emg,Chen:2021dsw}, see also Refs. \cite{Bowick:1985af,Susskind:1993ws,Horowitz:1996nw,Giveon:2006pr}.} This is the non-locality scale of the theory, of order the string length.

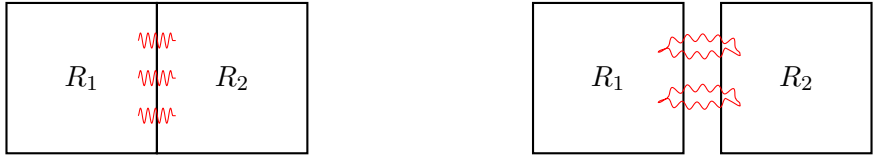
\begin{figure}
    \centering
    \begin{tikzpicture}
        \draw[thick] (0,0) -- (2,0) -- (2,2) -- (0,2) -- cycle;
        \draw[thick] (2,0) -- (4,0) -- (4,2) --(2,2)--cycle;
        \node[] at (1, 1) {$R_1$};
        \node[] at (3, 1) {$R_2$};
        \draw[red,decorate, decoration={snake, amplitude=1mm, segment length=1mm}] (1+3/4,1.5) -- (3-3/4,1.5);
        \draw[red,decorate, decoration={snake, amplitude=1mm, segment length=1mm}] (1+3/4,1) -- (3-3/4,1);
        \draw[red,decorate, decoration={snake, amplitude=1mm, segment length=1mm}] (1+3/4,0.5) -- (3-3/4,0.5);
        \draw[thick] (7,0) -- (9,0) -- (9,2) -- (7,2) -- cycle;
        \draw[thick] (9.5,0) -- (11.5,0) -- (11.5,2) -- (9.5,2) -- cycle;
        \draw[red,decorate, decoration={snake, amplitude=0.5mm, segment length=2mm}] (9.25,1.33) ellipse [x radius=0.5cm, y radius=0.125cm];
        \draw[red,decorate, decoration={snake, amplitude=0.5mm, segment length=2mm}] (9.25,0.66) ellipse [x radius=0.5cm, y radius=0.125cm];
        \node[] at (8, 1) {$R_1$};
        \node[] at (10.5, 1) {$R_2$};
    \end{tikzpicture}
    \caption{Left: High-energy modes in quantum field theory cause the entanglement between touching regions to be infinite. Right: In a theory of extended strings, the entanglement can diverge even at a finite distance.}
    \label{fig:divergences}
\end{figure}

A consequence of a Hagedorn spectrum on local algebras is that the split property breaks down at distances of order $\beta_H$. The split property has several equivalent formulations \cite{Doplicher:1984zz}, perhaps the most intuitive is that given two algebras $\mathcal{M}(R_1)$ and  $\mathcal{M}(R_2)$ associated to spacelike separated regions $R_1$ and $R_2$, the algebra generated by $\mathcal{M}(R_1)$ and $\mathcal{M}(R_2)$ is unitarily equivalent to the tensor product $\mathcal{M}(R_1) \otimes \mathcal{M}(R_2)$. This establishes the statistical independence of the two regions, there exist states in the Hilbert space that have no correlations between the two regions. An equivalent formulation that will be useful for us is that there exists a type I factor $\mathcal{N}$ such that
\begin{align}
    \mathcal{M}(R_1) \subset \mathcal{N}\subset \mathcal{M}(R_2)'
    \label{eq:split}
\end{align}
where $\mathcal{M}(R_2)'$ represents the commutant algebra, i.e.~all bounded operators on the Hilbert space that commute with $\mathcal{M}(R_2)$. In ordinary quantum field theory, the split property is expected to hold whenever $R_1$ and $R_2$ are not touching and when at least one of them is of finite volume, justifying lattice regularizations. When $R_1$ and $R_2$ touch, there is generically no tensor factorization due to the infinite number of high energy modes that entangle the two regions at the boundary. With this picture in mind, in theories with a Hagedorn temperature, such as string theory, the tensor factorization is broken even when $R_1$ and $R_2$ are a finite distance apart due to infinite entanglement generated by the string modes (see Fig. \ref{fig:divergences}). This has been rigorously established in free field theories
% In addition,
with a Hagedorn spectrum in Minkowski space \cite{DAntoni:1986olr} and argued for in 2D conformal field theories deformed by an irrelevant $T\bar{T}$ deformation \cite{Asrat:2020uib}.\footnote{See \cite{Folkestad:2023cze, Raju:2021lwh} for discussion of the breakdown of the split property in theories of gravity.}

In this paper, we investigate which structural properties of the theory can be inferred from the presence of a Hagedorn spectrum in the bulk string theory. We find that many familiar features remain intact, including the Reeh-Schlieder theorem \cite{reeh1961bemerkungen}, the existence of local operators, and the timelike tube theorem \cite{borchers1961vollstandigkeit,araki1963generalization}. However, in agreement with expectations from flat-space analyses, we demonstrate that the split property fails at distances of order $\beta_H$. In Section~\ref{sec:distalsplit}, we show explicitly that boundary regions of the holographic theory whose dual causal wedges are sufficiently close in the bulk violate the split property.

This breakdown of the split property in the bulk has interesting implications, which are explored in detail in Section~\ref{sec:impholo}. The statement is that disjoint regions in theories with holographic duals can violate the split property at large-$N$ and finite t'Hooft coupling. We emphasize that the violation of the split property arising from finite string tension is distinct from the superadditivity property studied in Ref.~\cite{Leutheusser:2022bgi,Leutheusser:2024yvf}, which also implies a different violation of the split property in the gravity regime.
We propose that the R\'enyi reflected entropy \cite{Dutta:2019gen} for R\'enyi index greater than one is a quantitative probe of the degree to which the split property is violated in the large-$N$ limit due to finite string tension effects. Furthermore, we show that the canonical purification of normal states associated with disjoint regions violating the split property leads to von Neumann algebras that are of type III. We comment on the potential conditions that different sub-types III emerge.
The appearance of a type III algebra that is not type III$_1$ would indicate that such regions are quantum-connected according to the algebraic ER=EPR conjecture of \cite{Engelhardt:2023xer}. 
% To the authors' knowledge, this construction constitutes the first appearance of a type III${}_{0}$ algebra in a setting of physical relevance.

\begin{figure}[t]
% \begin{align}
\centering
    \begin{tikzpicture}
        \draw[thick] (0,0) -- (4,4) -- (4,0) -- (0,4) --cycle ;
        \draw[decorate, decoration={zigzag}, thick] (0,0) -- (4,0);
        \draw[decorate, decoration={zigzag}, thick] (0,4) -- (4,4);
        \fill[gray, opacity = .25] (4,4)--(2,2) --(4,0)--cycle;
        \filldraw[red, opacity = .25] (2,2) -- (4,4) to[out = -135, in = 90] (3,2) to[out = -90, in = 135] (4,0)--cycle;
        % \node[red] at (2.5,2) {$\mathcal{A}_{sh}$};
        % \draw[very thick, blue] (4,0) -- (4,4);
        % \draw[very thick, blue,->] (4,0) -- (4,2);
    \end{tikzpicture}
    \hspace{3cm}
% \end{align}
% \centering
% \begin{align}\nonumber
    \begin{tikzpicture}
    \fill[gray, opacity = .25] ({2*cos(80)},{2*sin(80)}) arc[start angle=80, end angle=-80, radius=2] to[out = 105,in = -105] cycle;
    \fill[red, opacity = .25]  (0.3472, 1.969) to[out = -105,in = 105]  (0.3472, -1.969) to[out = 60, in = -60] (0.3472, 1.969);
    \draw[purple,thick] ({2*cos(80)},{2*sin(80)}) arc[start angle=80, end angle=-80, radius=2] to[out = 105,in = -105] cycle;
    \draw[thick] ({2*cos(80)},{2*sin(80)}) arc[start angle= 80, end angle= 440, radius=2];
    \end{tikzpicture}
% \end{align}
\caption{Left: The Penrose diagram of an AdS-Schwarzschild black hole with the stretched horizon of the right exterior region shaded in red. Right: A timeslice of AdS${}_{3}$, where the purple line is the Rindler horizon and the red region is the stretched horizon for the right region.
Any algebra with support in the red regions will not factorized from the white region, even though the regions are geometrically disjoint. }\label{fig:stretchedads3ex}
\end{figure}
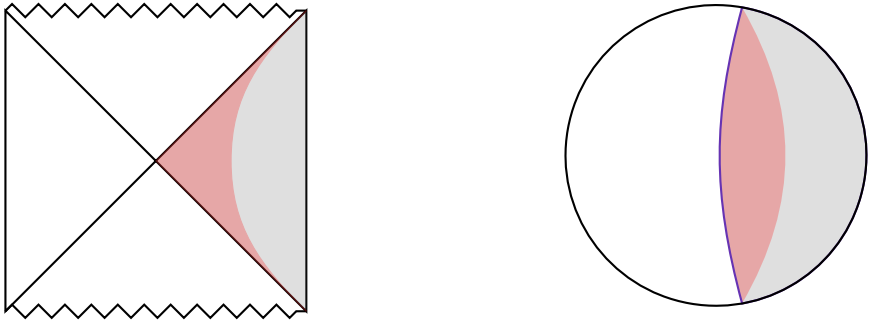

The failure of the split property is deeply connected to the physics of stretched horizons in string theory. Traditionally, the stretched horizon is defined as the region near a horizon where the locally measured Unruh temperature approaches the Hagedorn temperature, signaling the onset of strongly stringy physics. A sketch of the stretched horizon for an AdS black hole and time slice of a AdS-Rindler wedge in AdS${}_{3}$ is given in Fig. \ref{fig:stretchedads3ex}. The latter can be understood as a stretched Ryu-Takayanagi surface. The stretched horizon has been connected to a variety of phenomena, including the formation of a string gas near black hole horizons \cite{Susskind:1993ws,Hewitt:1993he} and longitudinal string spreading \cite{Karliner:1988hd,Susskind:1993aa}. These proposals suggest that the stretched horizon is not merely a convenient regulator but encodes fundamental aspects of gravitating horizons in string theory. In this work, we propose an algebraic definition of the stretched horizon, identifying it with the region where the split property fails, thereby signaling a breakdown in the factorization of the local operator algebra between the interior and exterior regions. From this perspective, the stretched horizon is not simply a spatial location but an intrinsically quantum-algebraic phenomenon. We make this correspondence precise by giving a formal definition of the stretched-horizon region in terms of the boundary algebra in Section \ref{sec:stringyhor}. 
In Section \ref{sec:discus}, we briefly explore the effects of turning on string interactions. 
Many of the results derived in the gravitational limit continue to hold at finite string tension, such as the crossed-product construction in Refs. \cite{Witten:2021unn, Chandrasekaran:2022eqq}. This framework allows us to rigorously define a notion of generalized entropy at finite, rather than merely infinitesimal, string tension. \\

\paragraph{Notation} Throughout the text, $a$, $b$, $\phi$, and $\mathcal{O}$ refer to operators, $\mathcal{A}$ and $\mathcal{M}$ correspond to $\star$ and von Neumann algebras respectively, while $\mathcal{B}$ refers to all bounded operators on Hilbert space, the superscript $'$ of an algebra refers to the commutant and $\delta$ denotes a geodesic distance, generically between two causally complete regions.

\section{Algebraic properties of free string field theory}
\label{sec:algprop2}

In this section, we explore the algebraic properties of free string field theory, treating it as an infinite collection of independent free quantum field theories.\footnote{In principle, tidal effects in non-maximally symmetric backgrounds can induce mixing between modes \cite{Tyukov:2017uig,Bena:2018mpb,Bena:2020iyw,Martinec:2020cml,Dodelson:2020lal,Engelsoy:2021fbk}. However, in the non-interacting limit, we expect that the action for excited string modes can always be diagonalized \cite{Gesteau:2024rpt}.} Finite collections of quantum field theories can easily be combined via a tensor product. Still, extra care must be taken to avoid pathologies that can occur when taking infinite tensor products. We will show that, when appropriately treated, the local algebras of free string field theory exhibit many of the desirable features of standard quantum field theories. In particular, they satisfy the axioms of algebraic quantum field theory, enabling us to prove the Reeh--Schlieder Theorem (Section~\ref{subsec:ReehSchlieder}), construct normalizable operators (Section~\ref{sec:normop}), and establish the timelike tube theorem (Section~\ref{subsec:timeliketube}).

\subsection{Algebras, Hilbert spaces, and all that}

For simplicity, in this section, we will consider only massive scalar quantum field theories, though the generalizations to higher-spin fields should be straightforward.  
Consider a free scalar field on a $d+1$ dimensional manifold with metric $g$ and mass $m$, $\phi_m(x)$. We can smear this field in spacetime with at ``test-function'' $f$ to obtain an operator
\begin{align}
    \phi_m(f) \equiv \int d^{d+1}x\sqrt{-g}f(x) \phi(x).
\end{align}
$f$ is taken to be smooth and have compact spacetime support so that $\phi_m(f)$ can be considered a (quasi-) local operator. A $\star$-algebra, $\mathcal{A}_m$, can be obtained by taking all finite sums and products of these operators, along with the identity, factored by the equations of motion $\phi_m((\Box_{\mathcal{M}}+m^2)f) = 0$ where $\Box$ is the d'Alembertian. The algebra is also subject to the covariant canonical commutation relations $[\phi_m(f_1),\phi_m(f_2)] = iE(f_1,f_2)$, where $E$ is the Pauli–Jordan commutator function.

Given a ``vacuum''\footnote{In general curved spacetimes, there is no distinguished state to call the vacuum, in which case $\omega_m$ represents an arbitrary ``physical'' state in the theory we wish to consider. Physical usually means Hadamard, i.e.~has the same short distance behavior as the vacuum state in Minkowski space.} state $\omega_m$, i.e.~a positive linear functional on $\mathcal{A}_m$ with $\omega_m(\mathbbm{1}) = 1$, we can build a Hilbert space, $\mathcal{H}_m$ by identifying each element of $\mathcal{A_m}$ with a vector and taking the Hilbert space completion with respect to the norm induced by $\omega_m$ (the GNS construction). Under the condition that $\omega_m$ is a Gaussian state, the resulting Hilbert space takes the form of a (separable) Fock space
\begin{align}
    \mathcal{H}_m = \bigoplus_{n = 0}^{\infty}\mathcal{H}_{m,1}^{\otimes_sn},
\end{align}
where $\otimes_s$ is the symmetrized tensor product and $\mathcal{H}_{m,1}$ is the single-particle Hilbert space generated by acting with $\phi(f)$ on, $\ket{\Omega_m}$, the vector representative of $\omega_m$ in $\mathcal{H}_m$.

When dealing with an infinite tower of massive fields, one might be tempted to take the infinite tensor product of the Hilbert spaces $\mathcal{H}_m$. However, this construction is too large (non-separable) to be physically meaningful. Instead, we select a reference state $\ket{\Omega_m}$ in each $\mathcal{H}_m$ and consider tensor-product states $\bigotimes_m \ket{\psi_m}$ in which $\ket{\psi_m} \neq \ket{\Omega_m}$ for only finitely many values of $m$. The Hilbert space $\mathcal{H}$ is then defined as the completion of this set of states.

It will furthermore be helpful to close the algebra in the weak operator topology to form a von Neumann algebra, $\mathcal{M}_m$. This is equivalent to taking the double commutant of the algebra $\mathcal{A}_m'' = \mathcal{M}_m$, where the commutant of an algebra is the set of bounded operators on the Hilbert space that commute with the algebra. For the global algebra that we constructed, only multiples of the identity operator commute with all elements, so $\mathcal{M}_m$ is the algebra of all bounded operators on the Hilbert space, $\mathcal{B}(\mathcal{H}_m)$. By definition, this is a type I von Neumann algebra. We may then consider von Neumann subalgebras of $\mathcal{B}(\mathcal{H}_m)$ that are supported only in local spacetime regions. In quantum field theory, these are universally type III$_1$ von Neumann algebras.

Just as when constructing the Hilbert space, we must be careful when taking the infinite tensor product of the above algebras. The global $\star$-algebra, $\mathcal{A}$, can be constructed by restricting to the subset of operators in $\bigotimes_m\mathcal{A}_m$ that only act nontrivially on a finite number of $m$'s. This algebra can subsequently be completed to a von Neumann algebra $\mathcal{M}$. For subalgebras restricted to a region $R$, we will refer to these algebras as $\mathcal{A}(R)$ and $\mathcal{M}(R)$ respectively.

\subsection{The Reeh-Schlieder theorem}
\label{subsec:ReehSchlieder}

The Reeh-Schlieder theorem states that the vacuum vector is cyclic and separating for local algebras, i.e.~if the set of vectors $a\ket{\Omega_m}$ for $a\in \mathcal{M}_m(R)$ is dense in the Hilbert space and $a \ket{\Omega_m } = 0$ for $a\in \mathcal{A}_m$, then $a = 0$. It is a general fact that if a state is cyclic for an algebra, then it is separating for its commutant. Thus, we only need to focus on the cyclicity property for local regions. The purpose of this section is to explain why nothing breaks down when the infinite tensor product of free fields is taken.

 It is known that the Reeh-Schlieder property holds for free massive scalar fields in curved spacetime \cite{Sanders:2008gs}. This means that for the algebras $\mathcal{M}_m(R)$, there exists a sequence $\{a_l\} \in \mathcal{M}_m(R)$ such that for every $\ket{\Psi_m} \in \mathcal{H}_m$
\begin{align}
    \lim_{l \rightarrow \infty} \parallel a_l \ket{\Omega_m} - \ket{\Psi_m}\parallel = 0.
\end{align}
Thus for every $\epsilon_m > 0$ there exists an $N_m$ such that for all $l> N_m$
\begin{align}
    \parallel a_l \ket{\Omega_m} - \ket{\Psi_m}\parallel < \epsilon_m.
\end{align}

When moving to the infinite tensor product, we may restrict to tensor product states because these form a basis for the entire Hilbert space. A general product state in $\mathcal{H}$ is of the form
\begin{align}
    \ket{\Psi} = \bigotimes_{n=1}^\infty \ket{\Psi_n}, \quad \sum_{n=1}^{\infty}\parallel \ket{\Omega_n} - \ket{\Psi_n}\parallel^2 < \infty.
\end{align}
For the sum to converge, there must exist an $N_{\Psi,\epsilon}$ such that 
\begin{align}
    \sum_{n= N_{\Psi,\epsilon}}^{\infty}\parallel \ket{\Omega_n} - \ket{\Psi_n}\parallel^2 < \epsilon 
\end{align}
for every $\epsilon >0$. We want to approximate $\ket{\Psi}$ by acting with an operator $a$ that has support on a finite number of mass sectors on the vacuum, thus in $\mathcal{M}(R)$. We ask that for every $\epsilon' >0$, there exists an $a$ such that
\begin{align}
    \frac{1}{2}\parallel a\ket{\Omega}-\ket{\Psi}  \parallel^2 = 1-\bra{\Psi}a\ket{\Omega} <  \epsilon',
\end{align}
where without loss of generality, we have assumed that $\bra{\Psi}a\ket{\Omega}$ is real.

For any fixed $\epsilon'$, we can take $\epsilon < \epsilon'/2$, which fixes a finite cutoff value $N_{\Psi,\epsilon}$. We have 
\begin{align}
    \bra{\Psi}a\ket{\Omega} &= \left(\prod_{n = 1}^{N_{\Psi,\epsilon}-1}\bra{\Psi_n}a_n\ket{\Omega_n}\right)\left(\prod_{n=N_{\Psi,\epsilon}}^{\infty} \bra{\Psi_n}\Omega_n\rangle\right)
    \\
    &
    \approx\left(\prod_{n = 1}^{N_{\Psi,\epsilon}-1}\bra{\Psi_n}a_n\ket{\Omega_n}\right)e^{-\frac{1}{2}\sum_{N_{\Psi,\epsilon}}^{\infty}\parallel \ket{\Psi_n} -\ket{\Omega_n}\parallel^2}
    % \\
    % &\geq e^{-\frac{1}{2}\sum_{n = 1}^{N_{\Psi,\epsilon}-1}\parallel \ket{\Psi_n} -a_n\ket{\Omega_n}\parallel^2}e^{-\frac{\epsilon}{2}},
\end{align}
where in the second line the approximation is valid because each $\bra{\Psi_n}\Omega_n\rangle$ in the second product is small.
By the Reeh-Schleider property for each mass sector, we can choose an $a_n$ such that
\begin{align}
    \bra{\Psi}a\ket{\Omega} &\geq e^{-\frac{1}{2}(\epsilon+\sum_{n = 1}^{N_{\Psi,\epsilon}-1}\epsilon_n)}\geq e^{-\frac{1}{2}(\epsilon'/2+\sum_{n = 1}^{N_{\Psi,\epsilon}-1}\epsilon_n)}
\end{align}
 Finally, we take $\epsilon_n =  \tfrac{\epsilon'}{2N_{\Psi,\epsilon}}$ which is possible because the Reeh-Schlieder property for each mass sector is independent of $\epsilon$ or $\epsilon'$. Thus, we have confirmed the cyclicity of vacuum for $\mathcal{M}(R)$.

 Given that the Reeh-Schlieder theorem holds, many results of Tomita and Takesaki's modular theory \cite{Borchers:2000pv} trivially generalize to theories with a Hagedorn spectrum, such as the Bisognano-Wichmann theorem \cite{Bisognano:1976za}. 
 These properties are necessary for defining stringy generalized entropy.

\subsection{Good operators}\label{sec:normop}

In this section, we consider some operators that arise in the completion of the algebra that have support on an infinite number of mass sectors. To get a ``good operator,'' meaning that it maps Hilbert space states to Hilbert space states, higher mass contributions must be sufficiently suppressed. We show how one can implement such suppression either by an explicit dampening factor or via smearing functions that are analytic in time.  

We first consider operators with $m$-dependent smearing functions that exponentially suppress high mass contributions, such as  
\begin{equation}\label{eq:explicitsup}
\mathcal{O}_{\textrm{con}}(x)=\sum_{m}e^{-Km/2}\phi_{m}(x) \ , \quad K>\beta_{H} \ .
\end{equation}
Due to the exponential suppression of high mass contributions, the two-point function of $\mathcal{O}_{\textrm{con}}(x)$ is convergent at any finite separation. For example, for any function $f$ with compact spacetime support, $\parallel \mathcal{O}_{\textrm{con}}(f)|\Omega\rangle\parallel <\infty$.  

A more interesting operator to consider is one where the higher mass contributions are not trivially suppressed, such as 
\begin{equation}\label{eq:undam}
\mathcal{O}_{\textrm{div}}(x)=\sum_{m}\phi_{m}(x) \ .
\end{equation}
As a result, the two-point function of $\mathcal{O}_{\textrm{div}}(x)$ diverges at sufficiently small but finite spacelike separations. Therefore, we cannot construct normalizable operators by spatial smearings of $\mathcal{O}_{\textrm{div}}(x)$. However, we can construct normalizable operators by appropriately smearing $\mathcal{O}_{\textrm{div}}(x)$ in time. Consider the smeared operator 
\begin{equation}
\mathcal{O}_{\textrm{div}}(f)=\sum_{m}\int d^{d}x f(x)\phi_{m}(x) 
\end{equation}
where we take $f(x)$ to be localized along a worldline with time coordinate $t$. We further take $f(x)$ to be analytic in $t$ in a complex strip of width $\alpha$. That $f(x)$ is analytic implies that it has support at all times (except for a measure zero set of times). We note that the maximum value of $\alpha$ that can be taken depends on the local temperature along the given worldline. Computing the contribution to the normalization from a single operator, we find 
\begin{equation}
\parallel\phi_{m}(f)|\Omega_{m}\rangle \parallel^{2}=\int \frac{d\omega}{2\pi} G_{m}(\omega)|\tilde{f}(\omega)|^{2}
\end{equation}
where 
\begin{equation}
\begin{split}
\langle \phi_{m}(t,\vec{x})\phi_{m}(t',\vec{x})\rangle&=\int \frac{d\omega}{2\pi} G_{m}(\omega) e^{-i\omega (t-t')}, \\
\tilde{f}(\omega)&=\int dt e^{i\omega t} \int d^{d-1}\vec{x} f(t,\vec{x})\sqrt{g} \ .
\end{split}
\end{equation}
%
% Assuming that $f(t,\vec{x})$ is non-zero for all times, 
The analyticity properties of $f(t,\vec{x})$ in time imply it decays exponentially at large $\omega$:
\begin{equation}
\tilde{f}(\omega) \sim e^{-\alpha |\omega|/2} \ .
\end{equation}
At parametrically large masses, this indicates that the integral is dominated by $\omega \sim m$, because $G_{m}(\omega)$ only has support on $\omega>m$. We are therefore led to the estimate 
\begin{equation}\label{estimateindividual}
\parallel\phi_{m}(f)|\Omega_{m}\rangle \parallel^{2} \sim C(m) e^{-\alpha m}
\end{equation}
where $C(m)$ is subexponential in $m$. Plugging the estimate in \eqcref{estimateindividual} into 
\begin{equation}
\parallel\mathcal{O}_{\textrm{div}}(f)|\Omega_{m}\rangle \parallel^{2}=\sum_{m} \parallel\phi_{m}(f)|\Omega_{m}\rangle \parallel^{2} \ ,
\end{equation}
we find the sum is convergent if $\alpha>\beta_{H}$. Therefore, we can construct normalizable smearings of $\mathcal{O}_{\textrm{div}}$ along worldlines provided the local temperature remains below the Hagedorn temperature. This condition automatically excludes worldline smearings of undamped operators, such as \eqcref{eq:undam}, that are sufficiently close to the horizon.

\subsection{Holographic reconstruction and timelike tube theorem}
\label{subsec:timeliketube}

We now briefly consider holographic reconstruction and the timelike tube theorem. 

In theories without a Hagedorn spectrum, bulk reconstruction was addressed by Hamilton, Kabat, Lifshitz, and Lowe (HKLL) in the large-$N$, large $\lambda$ limit using the extrapolate dictionary and the bulk equations of motion \cite{Hamilton:2006az}. $\mathcal{O}$ is a portion of the asymptotic boundary and $\mathcal{E}(\mathcal{O})$ is its so-called causal wedge. We represent bulk and boundary operators as
\begin{equation}
\begin{split}
\mathcal{O}_{\textrm{bulk}}(f)&=\sum_{m} \int dz d^{d-1}x^{\mu}\sqrt{g} f_{m}(z,x^{\mu})\phi_{m}(z,x^{\mu}) \\
\mathcal{O}_{\textrm{boundary}}(f)&=\sum_{m}  \int d^{d-1}x^{\mu} F_{m}(z,x^{\mu})\Phi_{m}(x^{\mu}) \\
\end{split}
\end{equation}
where $\Phi_{m}(x^{\mu})$ is the bulk dual of $\phi_{m}$ and $z$ is the radial coordinate. Bulk reconstruction can be interpreted as a map from $f_{m}(z,x^{\mu})$ to $F_{m}(z,x^{\mu})$. Given that the HKLL map holds for each free theory, the validity of HKLL in the Hagedorn theory follows from the normalizability of the bulk operator. 

HKLL reconstruction falls into the larger class of reconstruction problems that are addressed by the timelike tube theorem. The timelike tube theorem states that given an algebra of operators associated to a generally causally incomplete open set $\mathcal{O}$, the double commutant is equivalent to that for its causal completion, $\mathcal{E}(\mathcal{O})$, i.e.~$\mathcal{A}(R)'' = \mathcal{A}(\mathcal{E}(\mathcal{O}))''$. Borchers proved the timelike tube theorem in Minkowski spacetime for free fields \cite{borchers1961vollstandigkeit} and has since been generalized to a large class of curved spacetimes \cite{Strohmaier:2000ib}. These proofs are similar to the HKLL construction in AdS and therefore should immediately extend to free theories with a Hagedorn spectrum. However, the proofs of timelike tube for interacting fields, such as \cite{Strohmaier:2023hhy,Strohmaier:2023opz}, are fundamentally different and rely on certain analyticity properties of correlators and states. We believe it is likely that the timelike tube theorem extends to interacting theories with a Hagedorn spectrum; however, a rigorous proof is beyond the scope of this paper.\footnote{Many features that are consistent in free theories cease to hold once interactions are introduced. For instance, free massless higher-spin fields ($s>2$) can exist in isolation, but they become inconsistent as soon as interactions are turned on \cite{McGady:2013sga}.}

\section{Distal split property}
\label{sec:distalsplit}

Local quantum field theories are generally expected to obey split property whenever there is a finite distance between the regions of interest. The split property quantifies the intuition that the algebras of disjoint regions can be treated as if they act on independent Hilbert spaces. Under very general assumptions,\footnote{See discussion in Chapter V.5.2 of Ref. \cite{Haag:1996hvx}.} the split property is equivalent to the existence of a normal state $\omega$ on $\mathcal{A}(R_{1}) \vee  \mathcal{A}(R_{2})$ such that 
\begin{equation}\label{eq:productstatereq}
\omega(ab)=\omega_{1}(a)\omega_{2}(b)
\end{equation}
for all $a \in \mathcal{A}(R_1)$, $b \in \mathcal{A}(R_2)$,
where $\omega_1$ and $\omega_2$ are normal states on $\mathcal{A}(R_1)$ and $\mathcal{A}(R_2)$, respectively. The most well-studied examples of the split property involve the algebras of ball-shaped regions of radius $r$, $\mathcal{B}_{r}$, and their complements. In particular, it is known that $\mathcal{M}(\mathcal{B}_{r})$ and $\mathcal{M}(\mathcal{B}_{r+\delta}')$ are split for any $\delta > 0$. A finite gap between the regions is essential for the split property to hold. Without such a separation, arbitrarily high-energy modes would obstruct the existence of the product state in Eq.~\eqcref{eq:productstatereq}.

We will show that theories with a Hagedorn spectrum do not satisfy the split property. Instead, we demonstrate that theories with a Hagedorn spectrum obey a distal split property, which asserts the existence of a local product state, as described in \eqcref{eq:productstatereq}, only when the two regions are sufficiently separated. The non-existence of the product state will mark a fundamental breakdown of locality. Although interactions still occur at local points, regions that are adequately close cannot be treated as residing in independent Hilbert spaces. To prove the distal split property, we need to prove 
\begin{enumerate}
\item The non-existence of a product state for sufficiently close regions. 
\item The existence of a product state for regions that are sufficiently far apart. 
\end{enumerate}
The proof of the non-existence of a product state is fairly straightforward using bounds on the normalization of the product state in terms of correlation functions. The more difficult proof is showing that there does exist a product state for regions that are sufficiently separated. To prove this condition, we explicitly construct a product state in a ``doubled theory'' for sufficiently separated regions. Notably, the upper and lower bounds we derive do not coincide with each other, so we were not able to compute the exact distance at which the split property fails for the regions we consider, only that it is $\beta_{H}$ times an $O(1)$ number. We comment on other promising proof strategies in Section \ref{sec:discus}.

\subsection{Review: Minkowski space}

\begin{figure}[t]
\centering
\begin{align*}
\begin{tikzpicture}
    \filldraw[red, opacity = .25] (2,0) -- (0,2) -- (-2,0) -- (0,-2) -- cycle;
    \filldraw[gray, opacity = .25] (5,2.5) -- (2.5,0) -- (5,-2.5) -- cycle;
    \filldraw[gray, opacity = .25] (-5,2.5) -- (-2.5,0) -- (-5,-2.5) -- cycle;
    \draw[
        thick
        ] (2,0) -- (0,2) -- (-2,0) -- (0,-2) -- cycle;
    \draw[
        thick
        ] (5,2.5) -- (2.5,0) -- (5,-2.5);
        \draw[
        thick
        ] (-5,2.5) -- (-2.5,0) -- (-5,-2.5);
    \draw[<->] (2,0) -- (2.5,0) node[midway, above] {$\delta$};
    \draw[<->] (-2,0) -- (-2.5,0) node[midway, above] {$\delta$};
    \draw[<->] (-2,2.2) -- (0,2.2) node[midway, above] {$r$};
    \draw[dashed] (0,2) -- (0,2.2);
    \draw[dashed] (-2,0) -- (-2,2.2);
    \node[] at (0,0) {$\mathcal{M}(\mathcal{B}_{r})$};
    \node[] at (4,0) {$\mathcal{M}(\mathcal{B}_{r+\delta}')$};
    \node[] at (-4,0) {$\mathcal{M}(\mathcal{B}_{r+\delta}')$};
    \draw[->] (0,3) -- (0,4);
    \draw[->] (0,3) -- (-1,3);
    \node[] at (-1-0.2,3) {$|\vec{x}|$};
    \node[] at (0,4+0.2) {$t$};
    \end{tikzpicture}
\end{align*}
\caption{A sketch of the relevant regions in flat space.}\label{sketchrelflat}
\end{figure}
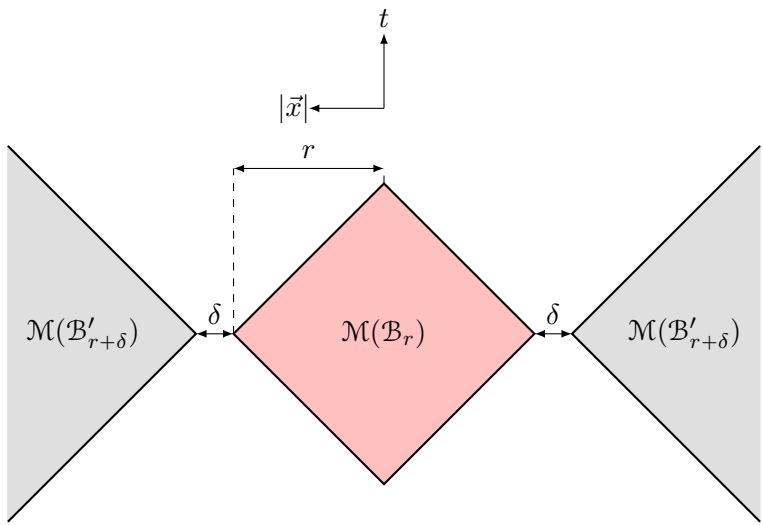

In this section, we first give proofs for the distal split property of ball-shaped regions, $\mathcal{B}_{r}$ and $\mathcal{B}_{r+\delta}'$ in Minkowski space in four spacetime dimensions for free theories with a Hagedorn spectrum. We sketch the relevant regions in Fig. \ref{sketchrelflat}. 
We will consider a tensor product of free scalar theories with the mass spectrum $m_{n}=\beta_{H}^{-1}\log(n)$ for $n\in \mathbb{Z}_{\geq 1}$. We denote the actual distance that the split property is restored as $\delta_{s}$. We first prove that the split property fails if $\delta\leq \beta_{H}/2$, which implies that $\delta_{s}>\beta_{H}/2$. We then then review the proof strategy in Ref. \cite{DAntoni:1986olr} that the split property holds if $\delta>\beta_{H}$, which implies that $\beta_{H}\geq \delta_{s}>\beta_{H}/2$. Again, these bounds are sub-optimal, but the proof strategy reviewed in this section generalizes to AdS.

\subsubsection{Proving $\delta_{s}>\beta_{H}/2$}\label{lowerboundds}

We now prove that the split property fails if $\delta$ is sufficiently small. Suppose we have an algebraic state $\omega_{\mathcal{B}_{r},\mathcal{B}_{r+\delta}'}$ that obeys the product constraint \eqcref{eq:productstatereq} for a theory with a Hagedorn spectrum. The existence of the product algebraic state implies there must be a ``product" vector in the Hilbert space that obeys 
\begin{equation}\label{eq:prodvectorder}
\langle \eta | ab'|\eta\rangle=\langle \eta | a|\eta\rangle \langle \eta | b'|\eta\rangle=\omega_{\mathcal{B}_{r},\mathcal{B}_{r+\delta}'}(ab')
\end{equation}
where $a\in \mathcal{M}(\mathcal{B}_{r})$ and $b'\in \mathcal{M}(\mathcal{B}_{r+\delta}')$. For $|\eta\rangle$ to be in the Hilbert space and not be the zero vector, it must have a non-zero inner product with at least one vector in the Hilbert space that is not itself: 
%`
\begin{equation}\label{infproductreq}
\exists |\Psi\rangle\in \mathcal{H} :\quad |\Psi\rangle \neq |\eta\rangle , \ \langle \Psi |\eta\rangle>0 \ .
\end{equation}
The goal is to show there is no such vector $|\eta\rangle$ such that \eqcref{infproductreq} is true when $\delta$ is sufficiently small. 

To simplify the problem, we take the product vector to be a tensor product over the product vectors of each free theory 
\begin{equation}\label{prodan}
|\eta\rangle= \bigotimes_{n=1}^{\infty} |\eta^{n}\rangle \ .
\end{equation}
The tensor product in \eqcref{prodan} is not the most general vector because none of the individual mass sectors are entangled with each other. However, we will now argue that the existence of a general vector satisfying Eq. (\ref{eq:prodvectorder}) implies the existence of a vector of the form Eq. (\ref{prodan}) that also satisfies Eq. (\ref{eq:prodvectorder}). In checking whether there is a state with $\langle \Omega |\eta\rangle>0$, we are attempting to find a vector with maximal fidelity $F(\ket{\Omega}\bra{\Omega}, \ket{\eta}\bra{\eta}) = |\bra{\Omega}\eta\rangle|^2$.
Suppose we had taken $\ket{\eta}$ not to be a tensor product as in \eqcref{prodan}. $\ket{\eta}\bra{\eta}$ would then have terms in in that mixed the different mass sectors.
We could then apply a quantum channel, $\mathcal{N}$, to $\ket{\Omega}\bra{\Omega}$ and $\ket{\eta}\bra{\eta}$ that kills the ``off-diagonal'' terms that mix mass sectors. This is a decoherence channel. $\mathcal{N}(\ket{\Omega}\bra{\Omega}) = \ket{\Omega}\bra{\Omega}$ because $\ket{\Omega}$ is a tensor product in the mass basis. However, $\mathcal{N}(\ket{\eta}\bra{\eta})$ becomes a mixed state $\sum_{i} p_i \ket{\eta_i}\bra{\eta_i}$ with $\sum_i p_i = 1$ where each $\ket{\eta_i}$ is a tensor product state between mass sectors. The fidelity between the resulting states, $\mathcal{N}(\ket{\Omega}\bra{\Omega})$ and $\mathcal{N}(\ket{\eta}\bra{\eta})$, is $\sum_i p_i  |\bra{\Omega}\eta_i\rangle|^2 \geq |\bra{\Omega}\eta\rangle|^2$, where the inequality comes from the monotonicity of fidelity under quantum channels. We may then take the $\ket{\eta_i}$ corresponding to the maximal $|\bra{\Omega}\eta_i\rangle|^2$ in the sum. $|\bra{\Omega}\eta_i\rangle|^2$ is larger than $\sum_i p_i  |\bra{\Omega}\eta_i\rangle|^2$, which was subsequently larger than $|\bra{\Omega}\eta\rangle|$. This justifies the use of the tensor product ansatz in \eqcref{prodan}
% . Therefore, it is sufficient to consider Eq. (\ref{prodan}) 
for the purpose of proving the non-existence of a product state satisfying \eqcref{infproductreq}.

We now argue that for the vector in Eq. (\ref{prodan}), satisfying the condition in Eq. (\ref{eq:prodvectorder}) implies that Eq. (\ref{infproductreq}) is violated, and therefore the vector lies outside the Hilbert space. From the infinite tensor product construction of the Hilbert space in section \ref{subsec:ReehSchlieder}, we know that every vector must approach the vacuum state when going to sufficiently high mass sectors. Therefore, for $\ket{\eta}$ to be in the Hilbert space, there will be some finite $N_{\eta}$, such that 
\begin{align}
    \prod_{n = N_{\eta}}^{\infty} |\bra{\eta^n}\Omega^n\rangle| > 0.
\end{align}
We prove that this infinite product
is identically zero for any finite $N_{\eta}$ when $\delta\leq \beta_{H}/2$ and so $\ket{\eta}$ cannot be in the Hilbert space. To see this, we note that the product being non-zero implies that 
\begin{equation}\label{boundfromsum}
\sum_{n=N_{\eta}}^{\infty} (1-|\langle \eta^{n}|\Omega^{n}\rangle|)<\infty
\end{equation}
which follows from $-\log(x)\geq 1-x$.
To show that \eqcref{boundfromsum} must diverge when $\delta<\beta_{H}/2$, we note that for pure states
\begin{equation}
\frac{1}{2}\parallel\omega_{\eta^{n}}-\omega_{\Omega^{n}} \parallel = \sqrt{1-|\bra{\Omega^n}\eta^n\rangle|^2}
\end{equation}
where 
\begin{equation}
\parallel \omega \parallel\equiv \sup_{\parallel A\parallel\leq 1} |\omega(A)| \ .
\end{equation}
Thus,
\begin{align}
     1-\sqrt{1-\frac{1}{4}|\bra{\Omega^n}A\ket{\Omega^n} - \bra{\eta^n}A\ket{\eta^n}|^2} \leq 1-|\bra{\Omega^n}\eta^n\rangle|
\end{align}
for any $A$ with operator norm less than one. 
We choose an operator of the form
\begin{equation}
A=\frac{\exp\left ( i\int d^{d-1}x d^{d-1}yf_{n}(x)g_{n}(y)\phi_{n}(x)\phi_{n}(y)\right )-1}{2}
\end{equation}
where $f(x)$ and $g(y)$ are smearing functions localized near the boundaries of $\mathcal{B}_{r}$ and $\mathcal{B}_{r+\delta}'$ respectively. $|\bra{\Omega^n}A\ket{\Omega^n} - \bra{\eta^n}A\ket{\eta^n}|$ is then related to the connected two-point function which decays exponentially in distance, so we find
\begin{equation}\label{lowerbound}
p(m_{n})e^{-2m_{n}\delta}\leq |1-\langle \Omega^{n}|\eta^{n}\rangle|
\end{equation}
where $p(m_{n})$ is some convergent polynomial in $m_{n}$. Plugging the lower bound in \eqcref{lowerbound} into the sum in \eqcref{boundfromsum}, it is clear that the sum diverges for $\delta\leq \beta_{H}/2$.

\subsubsection{Proving $\delta_{s}\leq \beta_{H}$}\label{upperboundds}

We now prove that the split property holds if the two regions are sufficiently separated, following the proof strategy in Ref. \cite{DAntoni:1986olr}.\footnote{AH thanks Nikita Sopenko for discussion of this proof.} Our goal is to explicitly construct a  normal state such that
\begin{equation}\label{innerprod}
a\in \mathcal{M}(\mathcal{B}_{r}), \ b\in \mathcal{M}(\mathcal{B}_{r+\delta}'):\quad \omega_{\mathcal{B}_{r},\mathcal{B}_{r+\delta}'}(ab)=\omega_{\mathcal{B}_{r},\mathcal{B}_{r+\delta}'}(a)\omega_{\mathcal{B}_{r},\mathcal{B}_{r+\delta}'}(b)
\end{equation}
when $\delta>\beta_{H}$. To explicitly construct such a normal state, we will consider a doubled theory where each mass sector now has two scalars with the same mass. We now have the algebraic tensor product of the von Neumann algebras $\mathcal{M}_{1}(\mathcal{B}_{r})$ and $\mathcal{M}_{2}(\mathcal{B}_{r})$ associated with the region $\mathcal{B}_{r}$ and a similar construction for region $\mathcal{B}_{r+\delta}'$. We will explicitly construct a normalizable product vector in this doubled theory, $|\eta_{t}\rangle\in \mathcal{H}_{1}\otimes \mathcal{H}_{2}$, such that 
\begin{equation}\label{innerprod2}
a\in \mathcal{M}_{1}(\mathcal{B}_{r}) , \ b\in \mathcal{M}_{1}(\mathcal{B}_{r+\delta}'):\quad \langle \eta_{t}|ab\otimes \mathbb{1}|\eta_{t}\rangle=\langle \eta_{t}|a\otimes \mathbb{1}|\eta_{t}\rangle \langle \eta_{t}|b\otimes \mathbb{1}|\eta_{t}\rangle \ ,
\end{equation}
Eq. (\ref{innerprod2}) can then be used to define the algebraic state in Eq. (\ref{innerprod}), where normalizability of $|\eta_{t}\rangle$ implies that the resulting product state, $\omega_{\mathcal{B}_{r},\mathcal{B}_{r+\delta}'}$, is well-defined. 

We initially consider a single mass sector, which now consists of two free scalars with the same mass. The vacuum of this theory is the tensor product 
\begin{equation}
|\Omega_{t}^{n}\rangle=|\Omega_{1}^{n}\rangle \otimes |\Omega_{2}^{n}\rangle \ ,
\end{equation}
where $|\Omega_{i}^{n}\rangle$ is the vacuum of theory $i$ at the $n^{th}$ mass level. Constructing a product vector in this doubled theory that obeys 
\begin{equation}\label{eq:doubledtheorybound}
2|1-\langle \Omega_{t}^{n}|\eta_{t}^{n}\rangle| \lesssim e^{-m_{n}\delta}  
\end{equation}
is sufficient to prove that 
\begin{equation}
0< \langle \eta_{t}|\Omega_{t}\rangle = \prod_{n=1}^{\infty} \langle \eta_{t}^{n}|\Omega_{t}^{n}\rangle  \ ,
\end{equation}
and so $\ket{\eta_t}$ is in the Hilbert space of the doubled theory.
Therefore, explicitly constructing a product vector in a single mass sector in the doubled theory that obeys \eqcref{eq:doubledtheorybound} is sufficient to show the existence of a vector obeying \eqcref{innerprod2}.   

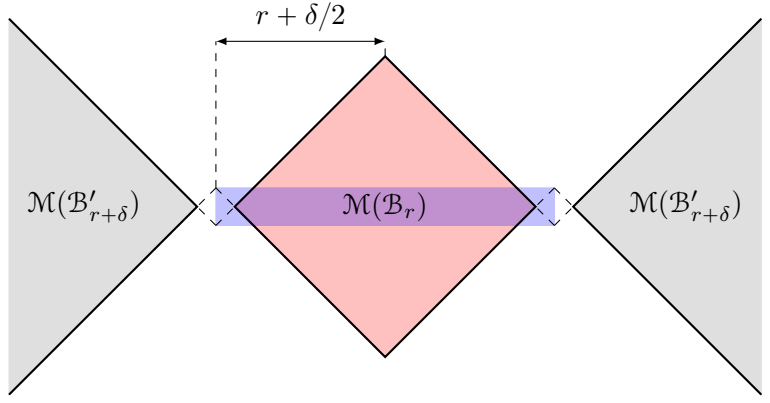
\begin{figure}[t]
\centering
\begin{align*}
\begin{tikzpicture}
    \filldraw[red, opacity = .25] (2,0) -- (0,2) -- (-2,0) -- (0,-2) -- cycle;
    \filldraw[blue, opacity = .25] (2.25,0.25) -- (2.25,-0.25) -- (-2.25,-0.25) -- (-2.25,0.25) -- cycle;
    \filldraw[gray, opacity = .25] (5,2.5) -- (2.5,0) -- (5,-2.5) -- cycle;
    \filldraw[gray, opacity = .25] (-5,2.5) -- (-2.5,0) -- (-5,-2.5) -- cycle;
    \draw[
        thick
        ] (2,0) -- (0,2) -- (-2,0) -- (0,-2) -- cycle;
    \draw[
        thick
        ] (5,2.5) -- (2.5,0) -- (5,-2.5);
        \draw[
        thick
        ] (-5,2.5) -- (-2.5,0) -- (-5,-2.5);
    \draw[dashed] (0,2) -- (0,2.2);
    \draw[dashed] (-2.25,0.25) -- (-2.25,2.2);
    \draw[dashed] (-2.25,-0.25) -- (-5,2.5);
    \draw[dashed] (-2.25,-0.25) -- (0,2);
    \draw[dashed] (2.25,-0.25) -- (5,2.5);
    \draw[dashed] (2.25,-0.25) -- (0,2);
    \draw[dashed] (-2.25,0.25) -- (-5,-2.5);
    \draw[dashed] (-2.25,0.25) -- (0,-2);
    \draw[dashed] (2.25,0.25) -- (5,-2.5);
    \draw[dashed] (2.25,0.25) -- (0,-2);
    \draw[<->] (-2.25,2.2) -- (0,2.2) node[midway, above] {$r+\delta/2$};
    \node[] at (0,0) {$\mathcal{M}(\mathcal{B}_{r})$};
    \node[] at (4,0) {$\mathcal{M}(\mathcal{B}_{r+\delta}')$};
    \node[] at (-4,0) {$\mathcal{M}(\mathcal{B}_{r+\delta}')$};
    \end{tikzpicture} \ .
\end{align*}
\caption{The blue region corresponds to where $j_{0}(f)$ has non-zero support in flat space.}\label{sketchsupportfl}
\end{figure}

We will construct an ansatz for $|\eta_{t}^{n}\rangle$ that is characterized by a single function. To do so, we leverage that the doubled theory has a global SO(2) symmetry with current 
\begin{equation}\label{explicitcurrent}
j_{\mu}=\phi_{1}\partial_{\mu}\phi_{2}-\phi_{2}\partial_{\mu}\phi_{1}
\end{equation}
which rotates the two flavors of scalars into each other. 
Consider the identity
\begin{equation}\label{desiredprop}
\exp\!\left( i \int d^{4}x \, f(x) j_{0} \right) \,
\phi_{1}(x) \,
\exp\!\left( - i \int d^{4}x \, f(x) j_{0} \right)
=
\begin{cases}
\phi_{2}(x), & x \in \mathcal{B}_{r}, \\[6pt]
\phi_{1}(x), & x \in \mathcal{B}_{r+\delta}', 
\end{cases}
\end{equation}
where $f(x)$ is chosen such that the above relation holds. A large class of functions $f(x)$ satisfying this identity is given by
\begin{equation}\label{ansatzfx}
f(x) = \frac{\pi}{\delta} \, \chi(\vec{x}) \, g\!\left( \tfrac{2x^{0}}{\delta} \right),
\end{equation}
where $\chi(\vec{x})$ is the characteristic function of a three-ball of radius $r+\delta/2$, and $g(x)$ satisfies
\begin{equation}\label{constraintsg}
g(x) = 0 \quad \text{for } |x|>1, 
\qquad \text{and} \qquad 
\int dx \, g(x) = 1 .
\end{equation}
The region where $f(x)$ has non-zero support is given in Fig. \ref{sketchsupportfl}.
To see that \eqcref{desiredprop} holds, we first consider $\phi_{1}(x)$ inserted in $\mathcal{B}_{r+\delta}'$. These operators commute with the current because they have disjoint support, so 
\begin{equation}\label{eqproof1}
\begin{split}
\forall x\in \mathcal{B}_{r+\delta}':\quad \exp \left ( ij_{0}(f) \right )\phi_{1}(x)\exp \left ( -ij_{0}(f) \right )=\phi_{1}(x)
\end{split}
\end{equation}
Now consider $\phi_{1}(x)$ in the region $\mathcal{B}_{r}$. These operators have the same commutation relations with the current as they would in the case where the current was integrated over all of space, so
\begin{equation}\label{equivalen}
\begin{split}
\forall x\in \mathcal{B}_{r}:\quad \exp \left ( ij_{0}(f) \right )\phi_{1}(x)\exp \left ( -ij_{0}(f) \right )=\exp \left ( ij_{0}(f_{b}) \right )\phi_{1}(x)\exp \left ( -ij_{0}(f_{b})\right )
\end{split}
\end{equation}
where
\begin{equation}
f_{b}(x)=\frac{\pi}{\delta}g(\frac{2x^{0}}{\delta}) \ .
\end{equation}
However, we can immediately identify 
\begin{equation}\label{jbdef}
j_{0}(f_{b})=\frac{\pi}{\delta}\int dx^{0} g(\frac{2x_{0}}{\delta}) Q=\frac{\pi}{2}Q
\end{equation}
where $Q$ is the conserved charge associated with the global SO(2) symmetry. Inserting \eqcref{jbdef} into \eqcref{equivalen}, one finds that 
\begin{equation}\label{eqproof2}
\forall x\in \mathcal{B}_{r}:\quad \exp \left ( ij_{0}(f) \right )\phi_{1}(x)\exp \left ( -ij_{0}(f) \right )=\phi_{2}(x) \ .
\end{equation}
Therefore, \eqcref{desiredprop} follows from Eqs. (\ref{eqproof1}) and (\ref{eqproof2}). We have not specified how the map in \eqcref{desiredprop} acts on operators outside of $\mathcal{B}_{r+\delta}'$ and $\mathcal{B}_{r}$, but it is irrelevant for our argument because we never consider these operators. 

Given Eq. (\ref{desiredprop}), we know that 
\begin{equation}
|\eta_{t}^{n}\rangle=e^{ij_{0}(f)}|\Omega_{t}^{n}\rangle
\end{equation}
where $|\eta_{t}^{n}\rangle$ obeys Eq. (\ref{innerprod2}). To show the bound in Eq. (\ref{eq:doubledtheorybound}) holds for some choice of $g(t)$, the relevant inner product is 
\begin{equation}
\parallel|\eta_{t}^{n}\rangle -|\Omega_{t}^{n}\rangle\parallel^{2}=\langle \Omega_{t}^{n}| (e^{-i j_{0}(f)}-1)(e^{i j_{0}(f)}-1)|\Omega_{t}^{n}\rangle
\end{equation}
but the right-hand side is hard to compute explicitly. However, using the inequality $(e^{-ix}-1)(e^{ix}-1)\leq x^{2}$, we find 
\begin{equation}\label{wantobound}
\langle \Omega_{t}^{n}| (e^{-i j_{0}(f)}-1)(e^{i j_{0}(f)}-1)|\Omega_{t}^{n}\rangle \leq \langle \Omega_{t}^{n}| j_{0}(f)^{2}|\Omega_{t}^{n}\rangle 
\end{equation}
where the right-hand side is much easier to compute using the explicit expression for the current in \eqcref{explicitcurrent}. To minimize \eqcref{wantobound}, we find it convenient to work in momentum space, so we define 
\begin{equation}\label{fourierF}
f(x)=\int \frac{d^{4}k}{(2\pi)^{4}} e^{-ikx}\tilde{f}(k) \ .
\end{equation}
and use the momentum space representation of the Wightman function
\begin{equation}\label{twopointprop}
\langle \Omega_{i}^{n}|\phi_{i}(x)\phi_{i}(y)|\Omega_{i}^{n}\rangle =\int \frac{d^{4}k}{(2\pi)^{3}}\theta(k^{0})\delta(k^{2}-m^{2})e^{-ik\cdot (x-y)} \ .
\end{equation}
The propagator is not time-ordered, which is why a delta function is used instead of the usual Feynman propagator. Inserting Eqs. (\ref{fourierF}) and (\ref{twopointprop}) into the right hand side of \eqcref{wantobound}, the inner product becomes 
\begin{equation}
\begin{split}
\langle \Omega_{t}| j_{0}(f)^{2}|\Omega_{t}\rangle &=\int \frac{d^{4}k}{(2\pi)^{4}} |\tilde{f}(k)|^{2} B(k)
\end{split}
\end{equation}
where
\begin{equation}
\begin{split}
B(k)&=\int \frac{d^{4}\ell}{(2\pi)^{2}} (k^{0}-2\ell^{0})^{2}\theta(k^{0}-\ell^{0})\delta((k-\ell)^{2}-m^{2})\theta(\ell^{0})\delta(\ell^{2}-m^{2}) \\
&=\frac{1}{12(2\pi)^{2}}\frac{\vec{k}^{2}(k^{2}-4m^{2})^{3/2}}{ (k^{2})^{3/2}}\theta(k^{2}-4m^{2})
\end{split}
\end{equation}
is the imaginary part of the time-ordered two-point bubble with some derivatives acting on the vertices. Making the change of variables $(k^{0})^{2}\rightarrow \kappa^{2}+\vec{k}^{2}$, the integral simplifies to 
\begin{equation}\label{minimizeinnerprod}
\begin{split}
\langle \Omega_{t}^{n}| j_{0}(f)^{2}|\Omega_{t}^{n}\rangle &=\frac{\pi^{2}}{48(2\pi)^{2}} \int_{4m^{2}}^{\infty} \frac{d\kappa^{2}}{2\pi}\int \frac{d^{3}\vec{k}}{(2\pi)^{3}} \frac{\left (1-\frac{4m^{2}}{\kappa^{2}} \right )^{3/2}}{\sqrt{\kappa^{2}+\vec{k}^{2}}}\vec{k}^{2}|\tilde{g}(\frac{\delta\sqrt{\kappa^{2}+\vec{k}^{2}}}{2})\tilde{\chi}(\vec{k})|^{2}
\end{split}
\end{equation}
which is a functional on $g(x)$. The goal is to find a choice of $g(x)$ obeying the constraints in \eqcref{constraintsg} that minimizes the inner product in \eqcref{minimizeinnerprod}.\footnote{We note that our formula for the inner product in \eqcref{minimizeinnerprod} differs slightly from Ref. \cite{DAntoni:1986olr}, which we think has some typos. However, both expressions give that $\delta_{s}\leq \beta_{H}$.}

To find an upper bound on the minimum of \eqcref{minimizeinnerprod}, we use that 
\begin{equation}
\forall \vec{k}: \quad |\tilde{\chi}(\vec{k})|\leq \frac{4 \pi }{3} (r+\delta/2)^{3}
\end{equation}
and that 
\begin{equation}
|\tilde{g}( \frac{\delta\sqrt{\kappa^{2}+\vec{k}^{2}}}{2})|^{2} \leq \frac{1}{(\kappa^{2}+\vec{k}^{2})^{3}} \left ( \frac{\delta}{2} \right )^{-6} \sup_{\omega \geq m\delta}|\omega^{3}\tilde{g}(\omega)|^{2} \ .
\end{equation}
From both these formulas, it follows that 
\begin{equation}
\langle \Omega_{t}^{n}| j_{0}(f)^{2}|\Omega_{t}^{n}\rangle \leq c \left ( \frac{4 \pi }{3}(r+\delta/2)^{3} \right )^{2}\left ( \frac{\delta}{2} \right )^{-6} \sup_{\omega \geq m\delta}|\omega^{3}\tilde{g}(\omega)|^{2} 
\end{equation}
where 
\begin{equation}\label{cintflat}
c\equiv \frac{\pi^{2}}{48(2\pi)^{2}} \int_{4m^{2}}^{\infty} \frac{d\kappa^{2}}{2\pi}\int \frac{d^{3}\vec{k}}{(2\pi)^{3}} \frac{\left (1-\frac{4m^{2}}{\kappa^{2}} \right )^{3/2}}{\sqrt{\kappa^{2}+\vec{k}^{2}}}\vec{k}^{2}(\kappa^{2}+\vec{k}^{2})^{-3} < \infty \ ,
\end{equation}
which is polynomial in $m$ and independent of $\delta$. Therefore, we have reduced the problem to minimizing $|\omega^{3}\tilde{g}(\omega)|^{2}$ under the constraints in \eqcref{constraintsg} and that $\omega>m\delta$. This is a purely mathematical problem, and its solution is presented in Appendix C of Ref. \cite{DAntoni:1986olr}:
\begin{equation}\label{upperboundgcom}
\inf_{g}\sup_{\omega>a} |\omega^{3}\tilde{g}(\omega)| \leq \textrm{constant}\ a^{3+\frac{1}{2}}e^{-\frac{a}{2}}
\end{equation}
where we impose that $g(x)$ obeys \eqcref{constraintsg}. The end result is that there exists a choice of $g(x)$ such that \eqcref{eq:doubledtheorybound} holds. Therefore, the split property holds when $\delta>\beta_{H}$. 

\subsection{AdS space}\label{adsdistal}

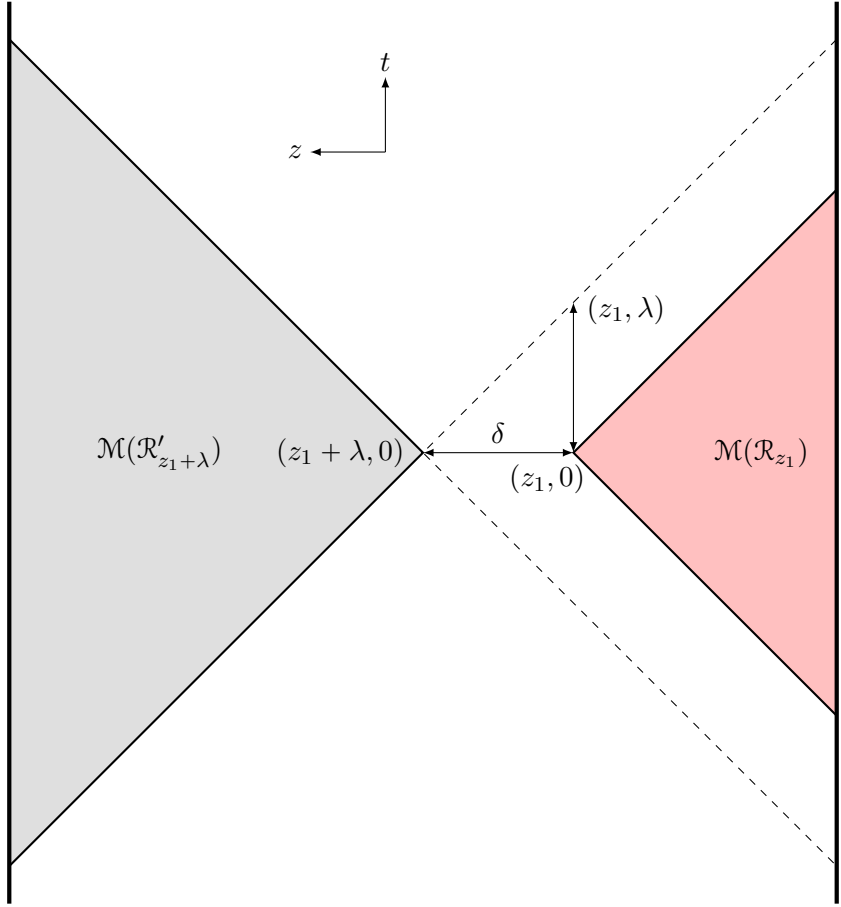
\begin{figure}[t]
\centering
\begin{align*}
\begin{tikzpicture}
    \filldraw[red, opacity = .25] (6,3.5) -- (2.5,0) -- (6,-3.5) -- cycle;
    \filldraw[gray, opacity = .25] (-5,5.5) -- (0.5,0) -- (-5,-5.5) -- cycle;
    \draw[
        thick
        ] (-5,5.5) -- (0.5,0) -- (-5,-5.5);
    \draw[
        ultra thick
        ] (6,-6) -- (6,6);
    \draw[
        ultra thick
        ] (-5,-6) -- (-5,6);
    \draw[
        thick
        ] (6,3.5) -- (2.5,0) -- (6,-3.5);
    \draw[
        dashed
        ] (6,5.5) -- (0.5,0) -- (6,-5.5);
    \node[] at (5,0) {$\mathcal{M}(\mathcal{R}_{z_{1}})$};
    \node[] at (-3,0) {$\mathcal{M}(\mathcal{R}_{z_{1}+\lambda}')$};
    \node[] at (2.5-0.35,-0.35) {$(z_{1},0)$};
    \node[] at (2.5+0.7,2-0.1) {$(z_{1},\lambda)$};
    \node[] at (0.5-1.1,0) {$(z_{1}+\lambda,0)$};
    \draw[<->] (0.5,0) -- (2.5,0) node[midway, above] {$\delta$};
    \draw[<->] (2.5,2) -- (2.5,0);
    \draw[->] (0,4) -- (0,5);
    \draw[->] (0,4) -- (-1,4);
    \node[] at (-1-0.2,4) {$z$};
    \node[] at (0,5+0.2) {$t$};
    \end{tikzpicture}
\end{align*}
\caption{A picture of the regions under considering in AdS${}_{2}$. $\delta$ is the geodesic distance between the relevant given points.}\label{ads2argpic}
\end{figure}

We now turn to the generalization to AdS. We consider AdS${}_{2}$ for simplicity, with the metric 
\begin{equation}
ds^{2}=\frac{dt^{2}-dz^{2}}{z^{2}}
\end{equation}
We consider the Rindler-AdS regions $\mathcal{R}_{z_{1}}$ and $\mathcal{R}_{z_{1}+\lambda}'$ in Fig. \ref{ads2argpic}. We will again consider a toy spectrum where 
\begin{equation}
\Delta_{n}=\frac{\log(n+1)}{\beta_{H}}, \quad \textrm{where}\quad m^{2}=\Delta_n(\Delta_n-1) \ .
\end{equation}
The only important ingredient is the existence of a Hagedorn temperature, so
the arguments generalize to theories with a more intricate spectrum, such as $\mathcal{N}=4$ super-Yang Mills. 

The argument that $\delta_{s}>\beta_{H}/2$ in Section \ref{lowerboundds} immediately generalizes to AdS because the behavior of two-point propagators at parametrically large mass is universal. It is given by 
\begin{equation}
\langle \phi(z,t)\phi(z',t') \rangle\sim e^{-m \delta} 
\end{equation}
where $\delta$ is the geodesic distance. For the regions in Fig. \ref{ads2argpic}, the relevant geodesic distance between points $(z_1,t)$ and $(z_1+\lambda,t)$
\begin{equation}\label{geodesiddi}
\delta=\log \left (\frac{z_{1}+\lambda}{z_{1}}\right ) \ ,
\end{equation}
where AdS units are implicit. It follows that the split property breaks if $\delta\leq \beta_{H}/2$, which implies that 
\begin{equation}
\delta_s > \frac{\beta_H}{2} ,\quad \lambda_{s}>z_{1}(e^{\frac{\beta_{H}}{2}}-1) \ .
\end{equation}
Therefore, it is clear that the split property breaks for causal wedges that are sufficiently close. 

The proof that the split property is restored at sufficiently large distances is much more difficult than the proof that the split property breaks at sufficiently short distances. The core difficulty is that propagators in curved space, even AdS, are significantly more complex than those in flat space, and even the large-mass approximation of the propagator in \eqcref{geodesiddi} is cumbersome. Additionally, we are considering AdS-Rindler regions with infinite volume. The proof strategy in the previous section breaks down for infinite volume regions in flat space due to infrared divergences. Although one should expect that such infrared divergences do not appear for Rindler-AdS regions,\footnote{A general rule of thumb is that putting a theory in AdS is analogous to putting it in a box  \cite{Alaverdian:2024llo}.} we must take care to show that such infrared divergences do not invalidate our calculation in intermediate steps.  

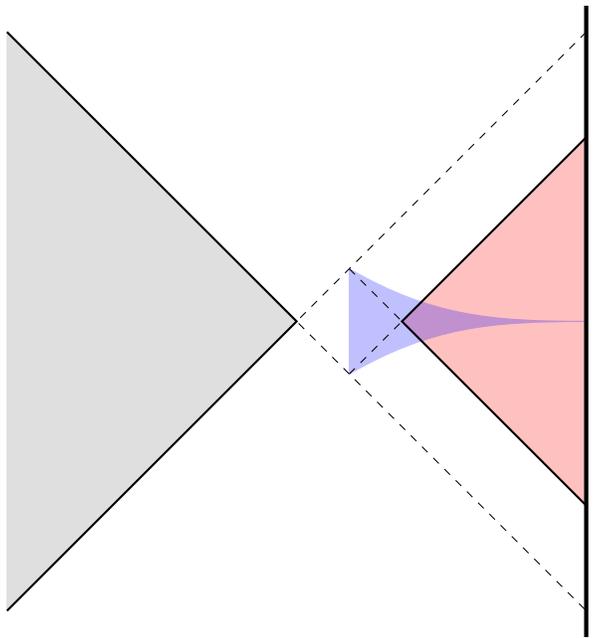
\begin{figure}[t]
\centering
\begin{align*}
\begin{tikzpicture}[scale=0.7]
    \filldraw[red, opacity = .25] (6,3.5) -- (2.5,0) -- (6,-3.5) -- cycle;
    \filldraw[blue, opacity = .25] (1.5,0) -- (1.5,1) to[out = -30, in = 180]  (6,0) to[out = 180, in = 30] (1.5,-1) -- cycle ;
    \filldraw[gray, opacity = .25] (-5,5.5) -- (0.5,0) -- (-5,-5.5) -- cycle;
    \draw[
        thick
        ] (-5,5.5) -- (0.5,0) -- (-5,-5.5);
    \draw[
        thick
        ] (6,3.5) -- (2.5,0) -- (6,-3.5);
    \draw[
        dashed
        ] (6,5.5) -- (0.5,0) -- (6,-5.5);
    \draw[
        dashed
        ] (1.5,-1) -- (6,3.5);
    \draw[
        dashed
        ] (1.5,1) -- (6,-3.5);
        \draw[
        ultra thick
        ] (6,-6) -- (6,6);
    \end{tikzpicture} \ . 
\end{align*}
\caption{The blue region is a sketch of where $j_{0}(f)$ has non-zero support.}\label{sketchj0ads}
\end{figure}

We nonetheless find that the proof strategy in Section \ref{upperboundds} remains viable with additional work. We follow the previous argument by considering a free scalar theory in $\textrm{AdS}_{2}$, doubling the theory and then defining the current for the resulting SO(2) global symmetry. However, our ansatz for $f(t,z)$ is slightly different than before
\begin{equation}\label{eq:ansatzads}
f(z,t)=\frac{\pi z}{\lambda'}g(\frac{2 t}{z \lambda'})\theta(z-z_{1}-\lambda/2)
\end{equation}
where $g(x)$ again obeys the constraints in \eqcref{constraintsg} and 
\begin{equation}
\lambda'=\frac{\lambda}{z_{1}+\lambda/2}
\end{equation}
A sketch of the region where $f(x)$ has support is given in Fig. \ref{sketchj0ads}.

We have again engineered $f(z,t)$ to have the desirable property that 
\begin{equation}
e^{ij_{0}(f)}\phi_{1}(z,t)e^{-ij_{0}(f)}=\begin{cases}
\phi_{2}(x), & x \in \mathcal{R}_{z_{1}}, \\[6pt]
\phi_{1}(x), & x \in \mathcal{R}_{z_{1}+\lambda}', 
\end{cases}
\end{equation}
so that the arguments in the previous section go through. The crucial formula is the AdS${}_{2}$ version of the charge Eq. (\ref{jbdef}):
\begin{equation}
Q=\frac{2}{\lambda'}\int \frac{dt\ dz}{z} g(\frac{2t}{z \lambda'})j_{0}(t,z) \ .
\end{equation}
Just as before, one finds that 
\begin{equation}
|\eta^n\rangle =e^{ij_{0}(f)} |\Omega_{t}^n\rangle
\end{equation}
induces a product state. The goal is again to minimize $\langle \Omega_{t}^n| j_{0}(f)^{2}|\Omega_{t}^n\rangle $, which is a double integral over the bubble with some derivatives, as a functional of $g(x)$. This is a conceptually straightforward problem. However, there is the technical issue that computing $\langle \Omega_{t}^n| j_{0}(f)^{2}|\Omega_{t}^n\rangle $ is now much more challenging because the AdS two-point bubble is much more complicated than its flat space counterpart. We therefore leave this computation to Appendix \ref{app:mintwopointbubble}, stating the final result:
\begin{equation}\label{endres}
\langle \Omega_{t}^{n}| j_{0}(f)^{2}|\Omega_{t}^{n}\rangle \lesssim (\ldots)\Delta_{n}^{-\Delta_{n}} +(\ldots) e^{-\lambda'\Delta_{n}}
\end{equation}
where the $(\ldots)$ corresponds to some functions of the relevant variables that are sub-exponential. The first term in \eqcref{endres} is so suppressed it is irrelevant for determining convergence. In contrast, the second term only converges when summing over mass sectors if $\lambda'>\beta_{H}$. Combining this result with the lower bound on the splitting distance, we have that 
\begin{equation}\label{boundondsAdS}
\log\left ( \frac{2+\beta_{H}}{2-\beta_{H}}\right )\geq \delta_{s}> \frac{\beta_{H}}{2}
\end{equation}
where $\lambda_{s}$ is the actual value of $\lambda$ at which the split property is restored. Note that $\lambda_{s}$ is not the geodesic distance between the two regions, which is instead given in Eq.~(\ref{geodesiddi}) as a function of $\lambda$ and $z_{1}$. Moreover, the upper bound in Eq.~(\ref{boundondsAdS}) is extremely weak, given that it actually diverges when $\beta_{H}=2$ in AdS units.

\begin{figure}[t]
\centering
\begin{align}\nonumber
    \begin{tikzpicture}
    \fill[gray, opacity = .25] ({2*cos(30)},{2*sin(30)}) arc[start angle=30, end angle=-30, radius=2] to[out = 150,in = -150] cycle;
    \draw[purple,thick] ({2*cos(30)},{2*sin(30)}) arc[start angle=30, end angle=-30, radius=2] to[out = 150,in = -150] cycle;
    \draw[thick] ({2*cos(80)},{2*sin(80)}) arc[start angle= 80, end angle= 440, radius=2];
    \fill[gray, opacity = .25] ({2*cos(30)},{2*sin(30)}) arc[start angle=30, end angle=-30, radius=2] -- ({1.2*cos(-30)},{1.2*sin(-30)})arc[start angle=-30, end angle=30, radius=1.2] -- cycle;
    \draw[red,thick]({2*cos(30)},{2*sin(30)}) arc[start angle=30, end angle=-30, radius=2] -- ({1.2*cos(-30)},{1.2*sin(-30)})arc[start angle=-30, end angle=30, radius=1.2] -- cycle;
    \draw[dashed] (0,0) circle[radius=1.2];
    \end{tikzpicture}
\end{align}
\caption{A timeslice of AdS${}_{3}$. The darker grey region, enclosed by the purple line, is the relevant patch of AdS${}_{3}$ that we are integrating over. Note that the bounds on the radial coordinate $z$ depend on $x^{\mu}$, which in this case is just the angular coordinate $\theta$. However, we can enlarge the integration domain to include the light grey region enclosed by the red line, so the integration bounds on $z$ are independent of the angular coordinate.}\label{fig:extendreg}
\end{figure}
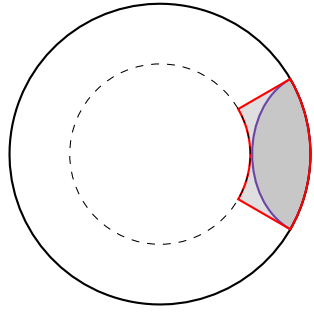

Finally, we restricted ourselves to AdS$_{2}$, but the same proof strategy applies to general dimensions. The only difficulty is that the upper bound on $z$ depends on $x^{\mu}$, which prevents us from immediately writing down an ansatz like \eqcref{eq:ansatzads}. A helpful trick is to upper bound the integral by enlarging the integration domain, as depicted in Fig. \ref{fig:extendreg}, so that the integral bounds on $z$ and $x^{\mu}$ factorize.

\subsection{Black Holes in AdS}

\begin{figure}[t]
\centering
\begin{align*}
    \begin{tikzpicture}
    \filldraw[red, opacity = .25] (4,0.5) -- (4,3.5) -- (2.5,2) -- cycle;
    \filldraw[black, opacity = .25] (0,0) -- (0,4) -- (2,2) -- cycle;
    \draw[
        thick
        ] (0,0) -- (4,4) -- (4,0) -- (0,4) --cycle ;
        \draw[decorate, decoration={zigzag}, thick] (0,0) -- (4,0);
        \draw[decorate, decoration={zigzag}, thick] (0,4) -- (4,4);
        \draw (4,0.5) -- (4.25,0.5);
        \draw (4,3.5) -- (4.25,3.5);
        \draw[<->] (4.25,0.5) -- (4.25,3.5) node[midway, right] {$T$};
        \draw[dashed] (2,2) -- (2,4.5);
        \draw[dashed] (2.5,2) -- (2.5,4.5);
        \draw[<->] (2,4.5) -- (2.5,4.5) node[midway, above] {$\delta$};
    \end{tikzpicture}
\end{align*}
\caption{A sketch of the relevant timeband algebra in an AdS-Schwarzchild background. $T$ is the temporal length of the timeband. If $\delta$ is sufficiently small, then the time band algebra is not split from the left algebra even though the regions are disconnected.}\label{fig:eternalAdSpic}
\end{figure}
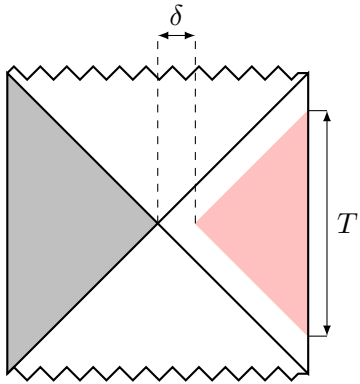

We could also consider regions connected by an eternal AdS-blackhole geometry, as in Fig. \ref{fig:eternalAdSpic}. Consider the AdS${}_{d+1}$-Schwarzchild metric 
\begin{equation}
ds^{2}=\frac{-f(z)dt^{2}+dz^{2}/f(z)+d\vec{x}^{2}}{z^{2}}, \quad f(z)=1-\frac{z^{d+1}}{z_{h}^{d+1}} \ .
\end{equation}
We note that our toy model of the spectrum as a tensor product of free fields with a Hagedorn spectrum is not quite sufficient to describe the bulk of string theory, even if we add spinning modes, in AdS${}_{d+1}$-Schwarzchild.\footnote{We thank Matthew Dodelson for discussion on this point.} The issue is that the excited string modes experience tidal forces in non-maximally symmetric spacetimes that lead to non-trivial effects, such as kinetic mixing, even in the non-interacting limit \cite{Tyukov:2017uig,Bena:2018mpb,Bena:2020iyw,Martinec:2020cml,Dodelson:2020lal,Engelsoy:2021fbk}. These corrections lead to non-trivial modifications of the bulk two-point functions. Nevertheless, we still expect a Hagedorn spectrum and that the two-point functions of parametrically heavy states exhibit the universal behavior discussed in section \ref{lowerboundds}.

Therefore, sticking with the toy spectrum considered previously, we can again show that such regions break the split property when the regions are sufficiently close using the universal behavior of two-point functions at large masses as in Section \ref{lowerboundds}. As a function of the length of the time-band, the geodesic distance of the causal wedge to the horizon, $\lambda$, is given by first solving for the radial depth of the timeband in the $z$ coordinate,
\begin{equation}
\int_{0}^{z} \frac{dz'}{1-\frac{z'^{d+1}}{z_{h}^{d+1}}}=\frac{T}{2} \ , 
\end{equation}
and then using that 
\begin{equation}
\delta=\int_{z}^{z_{h}}\frac{dz'}{z'\sqrt{1-\frac{z'^{d+1}}{z_{h}^{d+1}}}} \ .
\end{equation}
Timeband algebras will not be split with the left algebra when $\delta<\beta_{H}/2$. Such violations of the split property occur due to taking the large-$N$ limit in combination with finite $\lambda$. At finite-$N$, the two systems are obviously split. 

We believe it is very likely that the split property is restored for sufficiently small time-bands in the AdS-Schwarzchild background when $\delta$ is sufficiently large. However, we cannot consider an immediate generalization of the arguments in Section \ref{adsdistal} because the exact two-point bubble in an AdS-Schwarzchild background is not known in closed form and would likely receive non-trivial corrections from tidal forces. However, there is a good reason to believe that the split property is restored for sufficiently long distances. Since the geometry is asymptotically AdS, the calculation should again be free of infrared divergences as in $\textrm{AdS}_{2}$. The only issue is the UV behavior, which, again, should be largely insensitive to the details of the spacetime's large-scale structure. Therefore, we conjecture that the split property is restored for regions of $O(\beta_{H})$ distance from the horizon.

\section{Implications of Distal Split for Holography}\label{sec:impholo}

In this section, we discuss some interesting implications of the breakdown of the split property for AdS/CFT.

\subsection{Superadditivity}

As explored in Ref. \cite{Leutheusser:2022bgi,Leutheusser:2024yvf}, when taking the large-$N$ limit of the CFT, subregions on the boundary can correspond to different subregions of the bulk depending on the details of the limit. For a particular order of limits, even the gravity regime predicts violations of the split property that are distinct from those studied in Section \ref{sec:distalsplit}. 

\begin{figure}[t]
\centering
\begin{align}\nonumber
    \begin{tikzpicture}
    \fill[gray, opacity = .25] ({2*cos(60)},{2*sin(60)}) arc[start angle=60, end angle=-60, radius=2] to[out = 105,in = -105] cycle;
    \draw[purple,thick] ({2*cos(60)},{2*sin(60)}) arc[start angle=60, end angle=-60, radius=2] to[out = 105,in = -105] cycle;
    \fill[gray, opacity = .25] ({2*cos(-120)},{2*sin(-120)}) arc[start angle=-120, delta angle=-120, radius=2] to[out = -75,in = 75] cycle;
    \draw[purple,thick] ({2*cos(120)},{2*sin(120)}) arc[start angle=120, end angle=-120, radius=2] to[out = 75,in = -75] cycle;
    \draw[brown,thick] ({2*cos(120)},{2*sin(120)}) arc[start angle=120, end angle=60, radius=2] to[out = -105-35,in = -75+35] cycle;
    \draw[brown,thick] ({2*cos(-120)},{2*sin(-120)}) arc[start angle=-120, end angle=-60, radius=2] to[out = +105+35,in = 75-35] cycle;
    \node[] at (2.3,0) {$R_{2}$};
    \node[] at (-2.3,0) {$R_{1}$};
    \draw[thick] ({2*cos(80)},{2*sin(80)}) arc[start angle= 80, end angle= 440, radius=2];
    \end{tikzpicture}
\end{align}
\caption{A timeslice of AdS${}_{3}$. The two grey regions correspond to causal wedges of $R_{1}$ and $R_{2}$. The purple lines denote the minimal RT surfaces of the $R_{1}$ and $R_{2}$ regions individually. The brown lines together denote the RT surface of the union of the $R_{1}$ and $R_{2}$ regions.}\label{fig:newregion}
\end{figure}
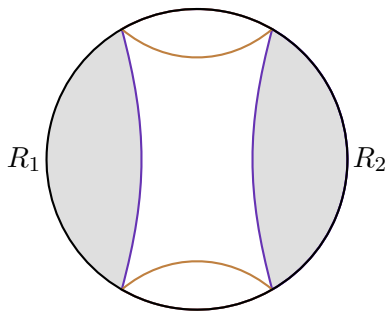

Consider the Hilbert space of the CFT on $S_{d}$ at integer $N$, $\mathcal{H}_N$, and the type I von Neumann algebra of all operators on the Hilbert space $\mathcal{B}(\mathcal{H}_N)$. We may associate type III algebras $\mathcal{M}_N(R_{1})$ and $\mathcal{M}_N(R_{2})$ to subregions $R_{1}$ and $R_{2}$ separated by some distance. When taking the large-$N$ limit, 
We work in the GNS Hilbert space obtained by acting on the vacuum with operators whose correlation functions remain of order one in the large-$N$ limit.\footnote{This definition may include operators that are not polynomials of single trace operators.} We may now define two, generally distinct, large-$N$ von Neumann algebras
\begin{align}
    \mathcal{M}_c(R_{1}\cup R_{2}) = \lim_{N\rightarrow \infty} (\mathcal{M}_N(R_{1})\vee \mathcal{M}_N( R_{2}))
    \\
    \mathcal{M}_d(R_{1}\cup R_{2}) = \left(\lim_{N\rightarrow \infty}  \mathcal{M}_N(R_{1}) \right) \vee \left(\lim_{N\rightarrow \infty} \mathcal{M}_N(R_{2})\right)
\end{align}
% where the subscript $c$ represents the causal complement.
$\mathcal{M}_d(R_{1}\cup R_{2})$ is contained in $\mathcal{M}_c(R_{1}\cup R_{2}) $. In the bulk, $\mathcal{M}_d(R_{1}\cup R_{2})$ corresponds to the disjoint union of the entanglement wedge algebras for $R_{1}$ and $R_{2}$. In contrast, $\mathcal{M}_c(R_{1}\cup R_{2})$ corresponds to the entanglement wedge algebra of $R_{1}\cup R_{2}$ which is larger if the entanglement wedge is ``connected,'' depicted in Fig. \ref{fig:newregion}. In most AdS/CFT literature, $\mathcal{M}_c(R_{1}\cup R_{2}) $ is the limit implicitly taken. 

We can quantify the degree to which the split property is broken by considering the R\'enyi reflected entropy, a quantity we now review. In quantum mechanics, when density matrices exist for local subregions, there is a canonical way to purify a mixed state by taking the square root of the matrix and interpreting that operator as a vector on a doubled Hilbert space $\ket{\sqrt{\rho_{R_1,R_2}}}_{R_1,R_2,R_1^*,R_2^*}$. The R\'enyi reflected entropy is the R\'enyi entropy of the subsystem $R_1 \cup \tilde{R}_1$ (equivalently the R\'enyi entropy of the subsystem $R_2 \cup \tilde{R}_2$)
\begin{align}
S_R^{(n)}(R_1,R_2) \equiv (1-n)^{-1}\log \Tr \rho_{R_1,\tilde{R}_1}^n.
\end{align}
Because the density matrix has unit trace, all R\'enyi entropies with $n > 1$ must be finite. When the algebras of $R_1$ and $R_2$ are type III, as in quantum field theory, there is a still a way to define reflected entropy as the entropy of a certain canonical type I factor \cite{Doplicher:1984zz} that appears in the split \eqcref{eq:split} \cite{Dutta:2019gen,Longo:2019pjj}. Therefore, if the split exists, the R\'enyi reflected entropy is finite. If the split does not exist, the R\'enyi reflected entropy will diverge.

As shown in Ref. \cite{Dutta:2019gen}, the reflected entropy for $n= 1$ has a very simple holographic dual, the area of the minimal cross-section of the entanglement wedge separating $R_1$ from $R_2$. When the entanglement wedge is in the connected phase, this is always $O(N^2)$ and so diverges in the large-$N$ limit. The R\'enyi reflected entropy has a similar holographic dual, though it is not quite as simple because one needs to account for backreaction between the replicas \cite{Dong:2016fnf}. Nevertheless, for $ n = 1+\epsilon$, this backreaction will be small and the R\'enyi reflected entropy can be seen to be $O(N^2)$. Therefore, there is a breakdown of the split property in the connected wedge phase. 

Therefore, we see that holography already predicts a drastic breakdown of the split property in the large-$N$ limit in the gravity regime. One result of this paper is that maintaining a finite string tension leads to a different source of violation. More concretely, the results of Section~\ref{sec:distalsplit} imply that even $\mathcal{M}_d(R_{1}\cup R_{2})$ fails to satisfy the split property when the bulk regions associated with $R_{1}$ and $R_{2}$ are sufficiently close. 
%In principal, this violation of the split property can occur even when $\mathcal{M}_{d}(R_{1}\cup R_{2})=\mathcal{M}_{c}(R_{1}\cup R_{2})$. 
To quantify the degree of split violation associated with $\mathcal{M}_{d}(R_{1}\cup R_{2})$, we can consider the reflected entropy of $\mathcal{M}_{d}(R_{1}\cup R_{2})$. That is, we first take large-$N$ and then compute the reflected entropy of the large-$N$ algebra. This is generically different than first computing the reflected entropy and afterwards taking the large-$N$ limit.

It is interesting to consider whether there exists a regime where $\mathcal{M}_{d}(R_{1}\cup R_{2})=\mathcal{M}_{c}(R_{1}\cup R_{2})$, but where the split property is nonetheless violated due to stringy excitations in the bulk. This may not always be possible because the Hagedorn temperature often has a finite lower bound in AdS units \cite{Sundborg:1999ue,Aharony:2003sx,Harmark:2018red}. Unfortunately, determining the minimum value of $\beta_{H}$ necessary for this scenario to occur lies beyond the scope of this paper, as we lack a precise determination of $\delta_{s}$.
% Nevertheless, we provide sufficient conditions for this scenario to arise in AdS${}_{3}$ in Appendix~\ref{minimalds}. 
The distance between the causal wedge of $R_1$ and $R_2$ in AdS$_3$ when the transition to a connected entanglement wedge occurs is $\log(3+2\sqrt{2})$ in AdS units \cite{Takayanagi:2017knl}.
It would be interesting if there were a general argument that the split property in the bulk cannot be violated via excited string modes if $\mathcal{M}_{d}(R_{1}\cup R_{2})=\mathcal{M}_{c}(R_{1}\cup R_{2})$. Such a restriction might lead to a lower bound on the Hagedorn temperature for generic large-$N$ theories.\footnote{We thank Juan Maldacena for discussion on this point.} In this setup, the reflected entropy should again diverge in the large-$N$ limit once the bulk regions become sufficiently close, reflecting the presence of stringy matter. At finite $N$, however, the split property is expected to be restored and the reflected entropy finite.

\subsection{Algebraic ER=EPR and an emergent type III algebra}\label{sec:erepr}

\begin{figure}[t]
\centering
\begin{align}\nonumber
    \begin{tikzpicture}
    \fill[gray, opacity = .25] ({2*cos(0)},{2*sin(0)}) arc[start angle=0, end angle=-85, radius=2] to[out = 95,in = 180] cycle;
    \draw[purple,thick] ({2*cos(0)},{2*sin(0)}) arc[start angle=0, end angle=-85, radius=2] to[out = 95,in = 180] cycle;
    \fill[gray, opacity = .25] ({2*cos(-95)},{2*sin(-95)}) arc[start angle=-95, delta angle=-85, radius=2] to[out = 0,in = 85] cycle;
    \draw[purple,thick] ({2*cos(-95)},{2*sin(-95)}) arc[start angle=-95, delta angle=-85, radius=2] to[out = 0,in = 85] cycle;
    \node[] at (1,-1) {$R_{2}$};
    \node[] at (-1,-1) {$R_{1}$};
    \draw[thick] ({2*cos(80)},{2*sin(80)}) arc[start angle= 80, end angle= 440, radius=2];
    \end{tikzpicture}
\end{align}
\caption{A timeslice of AdS${}_{3}$. The two grey regions correspond to $R_{1}$ and $R_{2}$. Even though the two regions are disconnected, we will assume that $\beta_{H}$ is sufficiently large that the split property is broken for $\mathcal{M}_{d}(R_{1}\cup R_{2})$.}\label{fig:newregion2}
\end{figure}
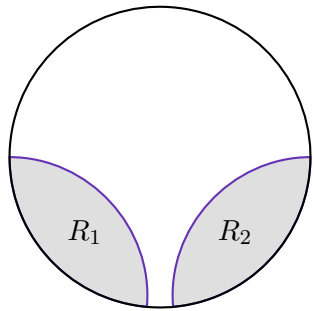

In this section, we investigate how stringy effects interact with the ER=EPR conjecture \cite{Maldacena:2013xja}, which states that when quantum systems are entangled, this is equivalent to them being connected by a wormholes. The ER=EPR conjecture was given a algebraic interpretation in Ref. \cite{Engelhardt:2023xer}.
Suppose we are given two entangled quantum systems, $R_{1}$ and $R_{2}$, which admit a well-defined $G \to 0$ limit. We consider the canonical purification of this state, which leads to a pure vector on the Hilbert space associated with the doubled system given by $R_{1}$, $\tilde{R}_{1}$, $R_{2}$, and $\tilde{R}_{2}$. We consider the joint algebra of systems $R_{1}\cup \tilde{R}_{1}$. The algebraic ER=EPR conjecture proposes:
\begin{itemize}
\item If $\mathcal{M}_{R_1\cup \tilde{R}_1}$ is type I, the original algebraic state corresponds to a disconnected geometry. 
\item If $\mathcal{M}_{R_1\cup \tilde{R}_1}$ is type III${}_{1}$, the original algebraic state corresponds to a classically connected geometry. 
\item Otherwise, the original algebraic state corresponds to a quantum-connected geometry. 
\end{itemize}
We will propose that geometrically disconnected regions violating the split property result in type III${}_{0}$ algebras, and are therefore quantum-connected, according to the above proposal. We consider two regions in AdS${}_{d+1}$ whose causal diamonds are geometrically disjoint but sufficiently close that the split property is violated. We refer to the systems associated with these regions as $R_{1}$ and $R_{2}$. See Fig. \ref{fig:newregion2} for a visual representation in AdS${}_{3}$. Finally, note that we are working with $\mathcal{M}_{d}(R_{1}\cup R_{2})$, not $\mathcal{M}_{c}(R_{1}\cup R_{2})$.

However, before considering the above construction, let us first analyze the Araki-Woods construction of von Neumann factors \cite{araki1968classification, powers1967representations}. Consider a system consisting of an infinite collection of entangled pairs of qubits. For each pair, we consider vectors the form
\begin{equation}
|\lambda\rangle = \sqrt{\frac{1}{1+\lambda}}\ket{0}_{R_1}\ket{0}_{R_2}+\sqrt{\frac{\lambda}{1+\lambda}}\ket{1}_{R_1}\ket{1}_{R_2},
\end{equation}
where $\lambda \in [0,1]$ parameterizes the degree of entanglement. For example, if $\lambda= 0$, the qubits are unentangled and if $\lambda = 1$, they are maximally entangled.
An infinite tensor product of such entangled states is
\begin{equation}
|\psi\rangle =  \bigotimes_{n=1}^{\infty}|{\lambda}_{n}\rangle 
\end{equation}
where the $\lambda_n$'s are generically distinct.
The infinite tensor product Hilbert space consists of acting on finite numbers of qubits on top of this state and taking the Hilbert space completion. The von Neumann algebras acting on a $R_1$
can be type I, II, or III depending on the sequence of ${\lambda}_{n}$'s. If the $\lambda_n$'s asymptotically approach $1$ sufficiently fast, then the algebra is type II$_1$. If they approach a value between $0$ and $1$, they are type III$_\lambda$. If they do not converge, the algebra is generically type III$_1$. If they approach $0$ (disentangle) sufficiently fast, then the algebra is type I. If they do not disentangle fast enough, the algebra can be type III$_0$. We refer the interested reader to Ref. \cite{araki1968classification} for the technical definition of ``fast enough.''

We can be somewhat more precise about the subtype classification. First, we define the modular operator for $R_1$ and state $\ket{\phi}$ in the infinite tensor product Hilbert space as $\Delta_{\phi}= S^{\dagger}_{\phi}S_{\phi}$ where $S_{\phi}$ is the antilinear operator defined by
\begin{align}
    S^{\dagger}_{\phi}a\ket{\phi} = a^{\dagger}\ket{\phi} ,
\end{align}
where $a$ is any operator associated to $R_1$. $\Delta_{\phi}$ is a positive operator. The Connes spectrum is then defined as the intersection of the spectra of $\Delta_{\phi}$ for all $\ket{\phi}$. If the Connes spectrum is $(0,\infty)$, the algebra is type III$_1$, if it is $\{\lambda^n\}$ for all integer $n$, it is type III$_{\lambda}$, if it is $\{1\}$, then the algebra is type III$_0$.

We now turn back to 
% arguing that the purified state leads to a type III${}_{0}$
% the algebra when restricted to 
the $R_{1}\cup \tilde{R}_{1}$ sub-algebra, using the same strategy as above, where the different entangled qubit systems will correspond to the distinct mass sectors in the stringy algebra. We denote the relevant systems for a single mass sector as $R_{i}^{n}$ and $\tilde{R}_{i}^{n}$.
The holographic background of the relevant Hilbert space for $R_{1}^{n} \cup \tilde{R}_{1}^{n}$ is constructed by CRT-reflecting the causal wedge across its extremal surface and gluing the reflected wedge back to the original along that surface \cite{Engelhardt:2018kcs}. 
An identical construction applies to $R_{2}$. For a single mass sector, the resulting algebra is of type I, reflecting the fact that the bulk regions associated with $R_{1}^{n}$ and $R_{2}^{n}$ are disconnected and not sufficiently entangled to be regarded as quantum connected. 
Nevertheless, because $R_{1}^{n}$ and $R_{2}^{n}$ are separated by a finite distance in the original geometry, the purified state for a single mass sector remains entangled between $R_{1}^{n} \cup \tilde{R}_{1}^{n}$ and $R_{2}^{n} \cup \tilde{R}_{2}^{n}$. The goal is to explain the conditions for this infinite tensor product of entangled vectors to lead to a various subtypes.

The first step is recognizing that the violation of the split property means that the algebra cannot be type I. If the algebra associated with $R_{1}\cup \tilde{R}_{1}$ were type I, then the split property would trivially hold. Therefore, the algebra is either type II or type III. However, the above tensor product of type I algebras should not yield a type II algebra because type II algebras are associated with tensor products of (near-)maximally entangled vectors. In contrast, the purified states of the higher-mass sectors become less entangled at higher masses because the correlation length decays. The infinite tensor product over mass sectors should subsequently be a type III algebra.

We now propose plausible scenarios where the result algebra is type III$_{\lambda}$ or type III$_{1}$. If the correct answer is that it is type III$_{\lambda}$, then under the algebraic ER=EPR conjecture, this describes a quantum-connected geometry. If it is type III$_1$, then we are in the situation where it is a classically connected geometry according to the algebraic ER=EPR proposal, even though it is not apparent. Type III$_1$ algebras admit half-sided modular inclusions and it would be interesting to understand the nature of such an inclusion.

In order to determine the subtype, it is necessary to understand the detailed spectrum of the reduced density matrix $\rho_{R_{1}\cup \tilde{R}_{1}}$ i.e.~the ``reflected entropy spectrum.'' While there has been some work on this \cite{Bueno:2020fle, Bueno:2020vnx, Dutta:2022kge}, to our knowledge, it has not been properly analyzed in the regime of interest, where regions are separated by a proper distance, $d$, much larger than the correlation length $m^{-1}$. The natural expectation is that in this regime, the R\'enyi reflected entropies decay exponentially
\begin{align}
    S_n(\rho^{(m)}_{R_{1}\cup \tilde{R}_{1}})~ \propto ~ e^{-2 m d}.
\end{align}
which is consistent with a leading order spectrum 
   \begin{align}
        \text{Spec}(\rho_{R_{1}\cup \tilde{R}_{1}}^{(m)}) = \{1 - e^{-2md}\sum_ic_i, c_i e^{-2md} , 0 \}
\end{align}
where $c_i$ are some $O(1)$ coefficients. The spectrum of the modular operator on a single mass would then be
    \begin{align}
        \text{Spec}(\rho_{R_{1}\cup \tilde{R}_{1}}^{(m)} \otimes {\rho_{R_{2}\cup \tilde{R}_{2}}^{(m)}}^{-1}) = \{1,\frac{1}{c_i e^{-2md}},c_i e^{-2md},\frac{c_i}{c_j}\}
    \end{align}
    up to subleading corrections. Now, if we take the tensor product of different $m$ sectors, where $m$ takes integer values, the spectrum is 
    \begin{align}
        \text{Spec}(\bigotimes _m\rho_{R_{1}\cup \tilde{R}_{1}}^{(m)} \otimes {\rho_{R_{2}\cup \tilde{R}_{2}}^{(m)}}^{-1}) = \{\left(\frac{c_i}{c_j}\right)^{p} \left(e^{-2d}\right)^q\}, \quad p,q \in \mathbb{Z}
    \end{align}
    and this is true for all states in the Hilbert space, so this defines the Connes' spectrum.

    In the case that there was only one $c_i$, the spectrum is that of a type III$_{\lambda}$ factor with $\lambda = e^{-2d}$. In the case that there is more than one $c_i$ and at least one $\frac{c_i}{c_j}$ is incommensurate with $e^{-2d}$ (or another $\frac{c_i}{c_j}$), which is the generic scenario, then the spectrum is dense on $\mathbb{R}$ and hence the algebra would be type III$_1$. If instead, the $c_i$'s are $m$-dependent in such a way that no fixed $\lambda\in (0,1)$ can be built (even approximately) from ratios of $c_i$'s involving only large $m$, then the algebra is type III$_0$. It is clearly of interest to have a precise understanding of the reflected entropy spectrum in the large-$m$ limit.

% Furthermore, since the amount of entanglement in each mass sector goes to zero at large-$m$, it may be natural to expect that the algebra is type III${}_{0}$. However, a more detailed analysis of the entanglement spectrum of the canonically purified state is required to definitively rule out other type III factors. 

To our knowledge, type III${}_{0}$ and type III${}_{\lambda}$ algebras have not previously appeared in quantum systems of physical relevance. Traditionally, the most ubiquitous von~Neumann algebras encountered in high-energy theory are type III${}_{1}$ factors, which furnish the operator algebras of local regions in quantum field theory~\cite{Araki:1964lyc,Longo:1982zz,Fredenhagen:1984dc} and, more recently, type II algebras that describe the subalgebras of ``dressed'' observables in perturbative quantum gravity~\cite{Witten:2021unn,Chandrasekaran:2022eqq,Jensen:2023yxy,Kudler-Flam:2023qfl}. 
Hyperfinite type III${}_{0}$ factors are far less explored because, unlike the other types, they are not unique up to isomorphism. Rather, there are an uncountably infinite number of them, characterized by certain ergodic flows \cite{takesaki1977flow}. 

\section{Stringy horizons}\label{sec:stringyhor}

We are now in a position to define the stretched horizon associated with a von Neumann algebra $\mathcal{M}$. Here, $\mathcal{M}$ may correspond, for example, to the algebra of observables in the exterior of a black hole, to an observer in de Sitter spacetime, or any other abstract von Neumann algebra. Our proposal for the stretched horizon builds on the work of Gesteau and Liu \cite{Gesteau:2024rpt}, who proposed an algebraic definition of the existence of a horizon that holds even at finite string tension. Consider a von Neumann algebra $\mathcal{M}$ and von Neumann subalgebras $\mathcal{M}_i$, called time-band algebras, associated to proper subregions of the theory. $\mathcal{M}$ is said to have a horizon if for every time-band algebra is a proper subalgebra of $\mathcal{M}$. This definition was shown to be equivalent to a simple condition on the spectral function \cite{Gesteau:2024rpt}.

Under the condition that a horizon exists, we would like to characterize it further. For example, is there a subalgebra of $\mathcal{M}$ that corresponds the horizon operators? We answer this in the affirmative for stringy horizons where a stretched horizon has a nontrivial $\star$-algebra which can be characterized by the breaking of the split. 

First consider the collection of von Neumann sub-algebras of $\mathcal{M}$ that are split with respect to the other algebra, $\mathcal{M}'$. We further restrict to the set of extremal minimally split algebras, which we denote as $S_{\mathcal{M}}$. To define $S_{\mathcal{M}}$, we first consider the set of von Neumann subalgebras of $\mathcal{M}$ that are minimally split. The minimally split algebras are von Neumann algebras that are not split from $\mathcal{M}'$, but are a limit of a sequence of subalgebras that are split from $\mathcal{M}'$
\begin{equation}
\ldots \supset \mathcal{M}_{i+2} \supset  \mathcal{M}_{i+1} \supset \mathcal{M}_{i} \ .
\end{equation}
The limit, $\mathcal{M}_{\infty}$, of the sequence is the smallest algebra containing all elements of the sequence as subalgebras. Next, we remove all minimally split algebras that are subalgebras of another minimally split algebra. The remaining von Neumann algebras are the extremal minimally split algebras. For every extremal minimally split algebra, we take the union with $\mathcal{M}'$ and compute the commutant (the relative commutant). The stretched horizon corresponds to the maximal $*$-union of these relative commutants, which we denote as $\mathcal{A}_{sh}$. As will soon be clear, we should not complete this to a von Neumann algebra because the result would be equal to $\mathcal{M}$ via the timelike tube theorem.

It is useful to work out a few examples explicitly to gain intuition for why this is a good definition of the stretched horizon algebra. We begin with the AdS-Schwarzschild black hole with $\mathcal{M}$, the algebra of the entire right exterior. According to the timelike tube theorem, this is equivalent to the von Neumann algebra associated with the right asymptotic boundary, which has a dual description as generated by single-trace operators in the gauge theory. This is a type III$_1$ von Neumann algebra and not split from its commutant $\mathcal{M}'$
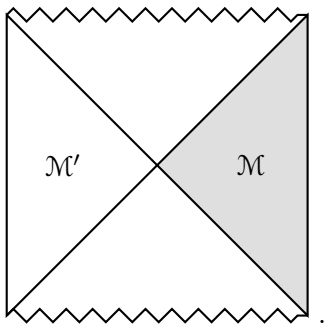
\begin{align}
    \begin{tikzpicture}
    \filldraw[gray, opacity = .25] (4,0) -- (4,4) -- (2,2) --cycle;
    \node[] at (3.25,2) {$\mathcal{M}$};
    \node[] at (.75,2) {$\mathcal{M}'$};
        \draw[
        thick
        ] (0,0) -- (4,4) -- (4,0) -- (0,4) --cycle ;
        \draw[decorate, decoration={zigzag}, thick] (0,0) -- (4,0);
        \draw[decorate, decoration={zigzag}, thick] (0,4) -- (4,4);
    \end{tikzpicture}\label{eq:adssch} \ .
\end{align}
We will demonstrate that the proposed definition of the stretched-horizon algebra indeed yields the desired algebraic structure in this context. We note that any von Neumann algebra associated with a causally complete region in the bulk can be defined in terms of the boundary algebra via the holographic dictionary \cite{Leutheusser:2022bgi,Engelhardt:2025bxy}. These generally are not associated with causally complete regions on the boundary.

We first consider the case where the string tension is taken to be infinitesimally small compared to the AdS scale. Hence, the QFT in curved space limit is valid, and the stretched horizon algebra should be trivial. In this case, the minimally split algebras are the set of all von Neumman sub-algebras of $\mathcal{M}$ with support on either the past or future horizon
\begin{align}
    \begin{tikzpicture}
        \filldraw[red, opacity = .25] (4,2) -- (4,4) -- (3,3) --cycle;
        \draw[
        thick
        ] (0,0) -- (4,4) -- (4,0) -- (0,4) --cycle ;
        \draw[decorate, decoration={zigzag}, thick] (0,0) -- (4,0);
        \draw[decorate, decoration={zigzag}, thick] (0,4) -- (4,4);
    \end{tikzpicture}
\end{align}
However, the final requirement that no element of $S_{\mathcal{M}}$ can be a sub-algebra of another element of $S_{\mathcal{M}}$ means that the only element of $S_{\mathcal{M}}$ is $\mathcal{M}$ itself. The relative commutant, and thus stretched horizon algebra is thus trivial, as expected when the split property is not violated. 

We progress to finite string tension, where we have understood that the split property breaks down at the string scale. The set of all algebras that are split with $\mathcal{M}'$ is all algebras which are localized at least an $\delta_{s}$ distance from the horizon, such as  
\begin{align}
    \begin{tikzpicture}
        \draw[
        thick
        ] (0,0) -- (4,4) -- (4,0) -- (0,4) --cycle ;
        \draw[decorate, decoration={zigzag}, thick] (0,0) -- (4,0);
        \draw[decorate, decoration={zigzag}, thick] (0,4) -- (4,4);
        \draw[blue,thick] (2,2) -- (4,4) to[out = -135, in = 90] (3,2) to[out = -90, in = 135] (4,0)--cycle;
        \draw[thick] (0,0) -- (4,4);
        \draw[thick] (4,0) -- (0,4);
        \filldraw[red, opacity = .25] (4,0.5) -- (4-0.6,0.5+0.6) -- (4,0.5+2*0.6) --cycle;
    \end{tikzpicture}
\end{align}
where the blue line denotes the points a $\delta_{s}$ distance from the horizon. The minimally split algebras are those whose tips are a $\delta_{s}$ distance from the horizon, such as
\begin{align}\label{validsub}
    \begin{tikzpicture}
        \draw[
        thick
        ] (0,0) -- (4,4) -- (4,0) -- (0,4) --cycle ;
        \draw[decorate, decoration={zigzag}, thick] (0,0) -- (4,0);
        \draw[decorate, decoration={zigzag}, thick] (0,4) -- (4,4);
        \draw[blue,thick] (2,2) -- (4,4) to[out = -135, in = 90] (3,2) to[out = -90, in = 135] (4,0)--cycle;
        \filldraw[red, opacity = .25] (4,0.5) -- (4-0.93,0.5+0.93) -- (4,0.5+2*0.93) --cycle;
        \draw[thick] (0,0) -- (4,4);
        \draw[thick] (4,0) -- (0,4);
    \end{tikzpicture}
\end{align}
These algebras are limits of sequences of algebras that do obey the split property. Finally, we remove all minimally split algebras that are subalgebras of another minimally split algebra, which leaves $S_{\mathcal{M}}$. An example of an element of $S_{\mathcal{M}}$ is \eqcref{validsub} because it is not a subalgebra of any other element of $S_{\mathcal{M}}$. 
The relative commutants correspond to bulk regions such as
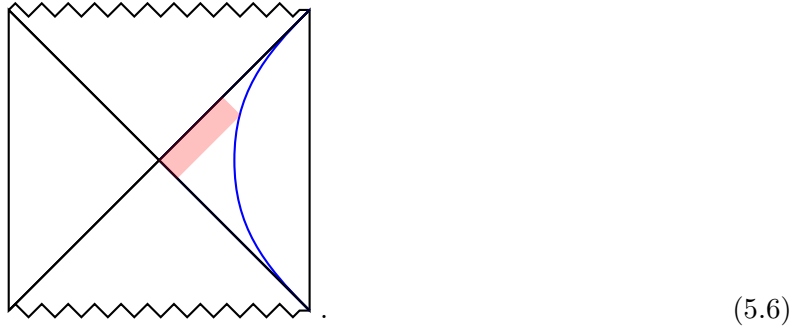
\begin{align}
\begin{tikzpicture}
        \draw[
        thick
        ] (0,0) -- (4,4) -- (4,0) -- (0,4) --cycle ;
        \draw[decorate, decoration={zigzag}, thick] (0,0) -- (4,0);
        \draw[decorate, decoration={zigzag}, thick] (0,4) -- (4,4);
        \draw[blue,thick] (2,2) -- (4,4) to[out = -135, in = 90] (3,2) to[out = -90, in = 135] (4,0)--cycle;
        \draw[thick] (0,0) -- (4,4);
        \draw[thick] (4,0) -- (0,4);
        \filldraw[red, opacity = .25] (2,2) -- (2.8375,2.8375) -- (3.075,2.6) -- (2.2375,2-0.2375) -- cycle;
    \end{tikzpicture} \ .
\end{align}
The $*$-union of such commutants has support along the entire stretched horizon region,
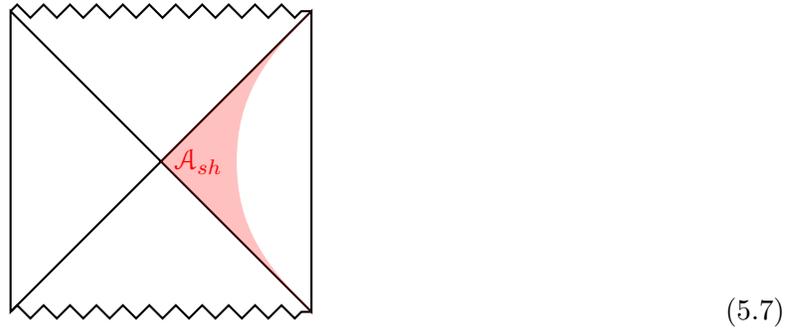
\begin{align}
    \begin{tikzpicture}
        \draw[
        thick
        ] (0,0) -- (4,4) -- (4,0) -- (0,4) --cycle ;
        \draw[decorate, decoration={zigzag}, thick] (0,0) -- (4,0);
        \draw[decorate, decoration={zigzag}, thick] (0,4) -- (4,4);
        \filldraw[red, opacity = .25] (2,2) -- (4,4) to[out = -135, in = 90] (3,2) to[out = -90, in = 135] (4,0)--cycle;
        \node[red] at (2.5,2) {$\mathcal{A}_{sh}$};
        % \draw[very thick, blue] (4,0) -- (4,4);
        % \draw[very thick, blue,->] (4,0) -- (4,2);
    \end{tikzpicture}
\end{align}
which is denoted in red. From this picture, it is clear that the von Neumann algebra completion of $\mathcal{A}_{sh}$ would yield $\mathcal{M}$ via the timelike tube theorem, so $\mathcal{A}_{sh}$ is, by construction, not a von Neumann algebra. 

We next consider the stringy generalization of the Ryu-Takayanagi surface. For simplicity, let us consider vacuum AdS${}_{3}$ with $\mathcal{M}$ corresponding to a boundary subregion
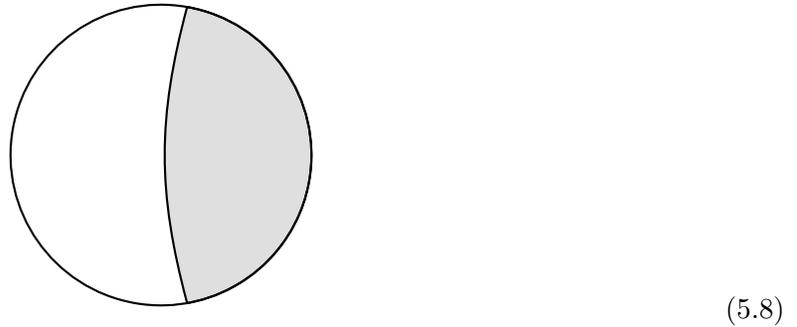
\begin{align}
    \begin{tikzpicture}
    \fill[gray, opacity = .25] ({2*cos(80)},{2*sin(80)}) arc[start angle=80, end angle=-80, radius=2] to[out = 105,in = -105] cycle;
    \draw[thick] ({2*cos(80)},{2*sin(80)}) arc[start angle=80, end angle=-80, radius=2] to[out = 105,in = -105] cycle;
    \draw[thick] ({2*cos(80)},{2*sin(80)}) arc[start angle= 80, end angle= 440, radius=2];
    \end{tikzpicture}
\end{align}
where we have drawn a timeslice of vacuum AdS. Applying the previous construction, the region where $\mathcal{A}_{sh}$ has non-trivial support is
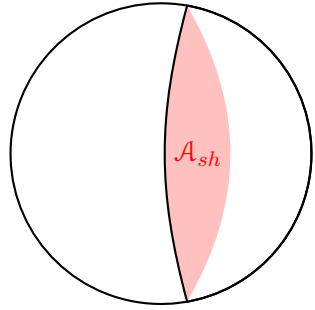
\begin{align}
    \begin{tikzpicture}
    \fill[red, opacity = .25]  (0.3472, 1.969) to[out = -105,in = 105]  (0.3472, -1.969) to[out = 60, in = -60] (0.3472, 1.969);
    \draw[thick] ({2*cos(80)},{2*sin(80)}) arc[start angle=80, end angle=-80, radius=2] to[out = 105,in = -105] cycle;
    \draw[thick] ({2*cos(80)},{2*sin(80)}) arc[start angle= 80, end angle= 440, radius=2];
      \node[red] at (.5,0) {$\mathcal{A}_{sh}$};
    \end{tikzpicture}
\end{align}
The region appears to disappear at the asymptotic boundary simply because the metric diverges there and the string length remains finite. We note that any boundary algebra localized in the right region that has support in the red region will not be split from the left algebra. 

This construction applies equally well to intervals of the observer wordlines in de Sitter
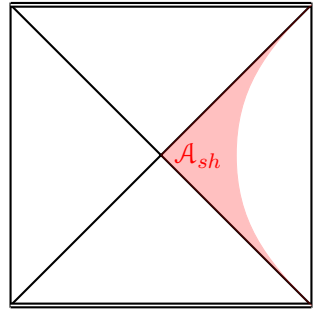
\begin{align}
    \begin{tikzpicture}
        \draw[
        thick
        ] (0,0) -- (4,4) -- (4,0) -- (0,4) --cycle ;
        \draw[double, thick] (0,0) -- (4,0);
        \draw[double, thick] (0,4) -- (4,4);
        \filldraw[red, opacity = .25] (2,2) -- (4,4) to[out = -135, in = 90] (3,2) to[out = -90, in = 135] (4,0)--cycle;
        \node[red] at (2.5,2) {$\mathcal{A}_{sh}$};
        % \draw[very thick, blue] (4,0) -- (4,4);
        % \draw[very thick, blue,->] (4,0) -- (4,2);
    \end{tikzpicture}
\end{align}
and Rindler horizons
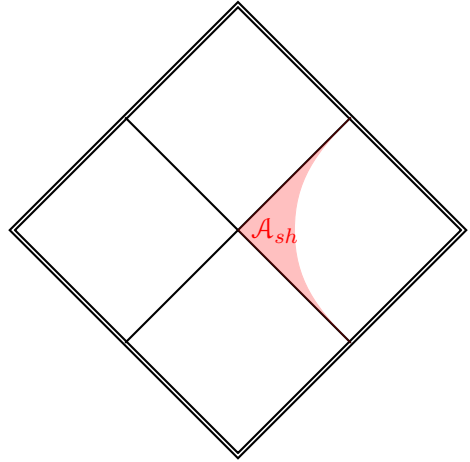
\begin{align}
    \begin{tikzpicture}
        \draw[thick,double] (0,0) --(3,3) -- (0,6)--(-3,3)--cycle;
        \draw[thick] (1.5,1.5) -- (-1.5,4.5);
         \draw[thick] (-1.5,1.5) -- (1.5,4.5);    \filldraw[red, opacity = .25] (0,3) -- (1.5,4.5) to[out = -135, in = 90] (.75,3) to[out = -90, in = 135] (1.5,1.5)--cycle;
                 \node[red] at (.5,3) {$\mathcal{A}_{sh}$};
    \end{tikzpicture}
\end{align}
and are therefore applicable beyond AdS/CFT contexts to the extent that string theory is well-defined on these target spaces.

\section{Discussion}\label{sec:discus}

In this paper, we considered the local operator algebra of holographic theories in the non-interacting limit at finite string tension. We demonstrated that such operator algebras can violate the split property and discussed several implications. For example, applying a canonical purification to the state of disjoint regions that violate the split property leads to a (conjectured) type III${}_{0}$ algebra, signaling how such regions are quantum connected. Using the apparent connection between the split property and the stretched horizon, we provided an algebraic definition of the stretched-horizon region in terms of the boundary algebra. There are numerous directions for follow-up work.

Throughout, we have considered the strict large-$N$ (small-$G$) limit, where the theory can be treated as a quantum field theory in curved spacetime without backreaction. It would be interesting to consider perturbative corrections. The most viable option is in the framework of AdS/CFT because of our ability to work with the algebra of observables perturbatively in $1/N$ for genuinely finite string tension by appealing to the boundary theory. In the supergravity regime, such corrections were shown to lead to a well-defined notion of von Neumann entropy, up to a state-independent constant \cite{Chandrasekaran:2022eqq}. The generalization to perturbative $\alpha'$ corrections was discussed in Ref. \cite{Kudler-Flam:2023qfl}, which showed the entropy was given by the generalized Wald entropy \cite{Wald:1993nt}. The technical method of the crossed product crucially relied on modular theory, which we showed in section \ref{sec:algprop2} still holds when there is a Hagedorn temperature. We do not know what to expect in the truly stringy regime, but this appears to be within reach.

From a technical perspective, it is desirable to explicitly compute $\delta_{s}$ in the noninteracting limit rather than relying solely on upper and lower bounds as in Section \ref{sec:distalsplit}. Our upper bound is particularly weak given that it diverges when $\beta_{H}=2$ in AdS units. Several proof strategies in the literature, such as those based on the nuclearity index, could in principle be adapted to yield a definitive value for $\delta_{s}$ \cite{Buchholz:1986bg,Buchholz:1991ie}.
The most natural value of $\delta_{s}$ corresponds to the point at which the local temperature reaches the Hagedorn temperature. Such a result would fit elegantly with the stretched horizon picture.

Recent progress has been made in better understanding holography at finite string tension and it would be interesting to understand the relation to our work. For instance, Ref. \cite{Chandrasekaran:2021tkb} analyzed holographic reconstruction in the small-tension regime using the eikonal approximation, whose stringy corrections are under perturbative control \cite{Shenker:2014cwa}. A similar calculation from a complementary perspective appeared in Ref. \cite{Caron-Huot:2022lff}. Moreover, the quantum error-correcting properties of SYK have also been investigated in this context \cite{Chandrasekaran:2022qmq}. The SYK model \cite{Maldacena:2016hyu} is conjectured to be dual to a finite-tension string theory, as its higher-spin bulk fields fail to decouple, see e.g.~\cite{Goel:2021wim,Vegh:2025kgx} for concrete proposals. 

Another interesting direction for future research is to gain a better understanding of the type of algebra that characterizes the quantum-connected geometries. Are these generic for stringy algebras or a feature of the regime that we work in? This requires a better understanding of the reflected entropy spectrum at asymptotically large masses.
% Our argument for the type III$_0$ factor was quick and there remains the possibility that the reflected entropy spectrum goes to zero in such a way that the algebra is a different type III factor.
%
If this leads to a type III$_0$ algebra, this would provide strong physical motivation for further study of III${}_{0}$ von Neumann algebras, which have previously been confined to mathematics.
We note that type III${}_{0}$ von Neumann algebras are deeply connected to ergodic theory \cite{connes1985approximately} and perhaps there is interesting interplay with the connections between emergent spacetime and the ergodic hierarchy discussed in Ref. \cite{Gesteau:2023rrx}.

Finally, it is natural to ask how our definition of the stretched horizon connects with other near-horizon phenomena, including longitudinal string spreading \cite{Karliner:1988hd,Susskind:1993aa}, the dynamics of a near-horizon string gas \cite{Susskind:1993ws,Hewitt:1993he}, and high-energy near-horizon scattering at finite string tension \cite{Shenker:2014cwa}. However, many of these phenomena are intrinsically tied to string interactions, which we largely ignored in this paper.\footnote{In addition, we note that interactions near horizons that probe stringy phenomena are 
typically characterized by 
energy and time scales that grow linearly with $N$. Such regimes lie somewhat 
outside the standard domain of von Neumann algebra techniques at large-$N$, 
which are usually formulated for observables whose energies, momenta, and time scales remain finite, neither parametrically small nor large, in the 
$N \to \infty$ limit. Recent work has begun to bridge this gap; see Ref.~\cite{Penington:2025hrc} for progress in extending algebraic methods to such high-energy/long-time phenomena.}
More recently, the stretched horizon has been conjectured to play a central role in proposed holographic duals of the double-scaled SYK model \cite{Susskind:2022bia,Rahman:2022jsf}. It would be interesting to understand whether the discussion in this paper has any quantitative implications for these conjectures.

\appendix

\paragraph{Acknowledgements} We thank Matthew Dodelson, Elliot Gesteau, Daniel Harlow, Sam Leutheusser, Juan Maldacena, Gautam Satishchandran, Urs Schreiber, Nikita Sopenko, Douglas Stanford, and Edward Witten for helpful discussions. AH is grateful to the Simons Foundation as well as the Edward and Kiyomi Baird Founders’
Circle Member Recognition for their support.  JKF is supported by the Marvin L. Goldberger Member Fund at the Institute for Advanced Study and the National Science Foundation under Grant No. PHY-2514611.

\appendix

\section{Minimizing the two-point function of the current}\label{app:mintwopointbubble}

We will initially consider an even simpler version of the problem by studying 
\begin{equation}\label{testexp}
B[g(x)]=\int\frac{dzdt_{1}}{z^{2}}\frac{dz'dt_{2}}{z'^{2}}f(z,t_{1})f(z',t_{2})\langle \phi(z,t_{1})\phi(z',t_{2})\rangle^{2}
\end{equation}
instead of $\langle \Omega_{t}^{n}| j_{0}(f)^{2}|\Omega_{t}^{n}\rangle $, which involves time derivatives acting on the two-point propagators in intermediate steps:
\begin{equation}\label{fullexpression}
\begin{split}
\langle \Omega_{t}^{n}| j_{0}(f)^{2}|\Omega_{t}^{n}\rangle=&\int\frac{dzdt_{1}}{z^{2}}\frac{dz'dt_{2}}{z'^{2}}f(z,t_{1})f(z',t_{2}) \\
&\times [(\partial_{t_{1}}-\partial_{t_{1}'})^{2}\langle \phi(z,t_{1})\phi(z',t_{2})\rangle\langle \phi(z',t_{1}')\phi(z'',t_{2}')\rangle]|_{t_{i}'\rightarrow t_{i},z_{i}'\rightarrow z_{i}} \ .
\end{split}
\end{equation}
The time-derivatives acting on the two-point propagators in \eqcref{fullexpression} do not significantly modify the computation strategy we apply to bound $B[g(x)]$. We will discuss how the following computation strategy is modified for \eqcref{fullexpression} compared to \eqcref{testexp} at the end of this subsection. Since we only work with a single mass sector, we will denote $\Delta_{n}$ and $\Delta$ for this section. 

To compute $B[g(x)]$ explicitly, we use the momentum space form of the bulk two-point Wightman function \cite{Liu:1998ty,Meltzer:2020qbr} 
\begin{equation}\label{momentumspaceAdS}
\langle \phi(z,t)\phi(z,t') \rangle=G_{\Delta}(t,z;t',z')=\int \frac{d\omega}{2\pi}e^{-i \omega (t-t')} \pi \sqrt{zz'} J_{\Delta-1/2}(\omega z)J_{\Delta-1/2}(\omega z')\theta(\omega) \ .
\end{equation}
and the product identity \cite{Fitzpatrick:2010zm,Fitzpatrick:2011dm,Dusedau:1985ue}
\begin{equation}\label{prodGform}
\begin{split}
G_{\Delta'}(t,z;t',z')^{2}&=\sum_{i=0}^{\infty} a_{\Delta'}(i)G_{2\Delta'+2i}(t,z;t',z') \\
a_{\Delta'}(i)&=\frac{(1/2)_{i}(2\Delta'+2i)_{1/2}(2\Delta'+i)_{1/2}}{2\sqrt{\pi}i! (\Delta'+i)^{2}_{1/2}(2\Delta'+i-1/2)_{n}}
\end{split}
\end{equation}
where $(x)_{i}$ is the Pochammer symbol. Using Eqs. (\ref{momentumspaceAdS}) and (\ref{prodGform}), we can explicitly compute $B[g(x)]$ as an integral sum
\begin{equation}\label{functionalint}
\begin{split}
B[g(x)]&=\sum_{i=0}^{\infty}a_{\Delta}(i) \frac{\pi^{3}\lambda'}{2}\int_{0}^{z+\lambda/2} \frac{dz}{z^{1/2}}\frac{dz'}{z'^{1/2}}\int \frac{d\omega_{1}d\omega_{2}}{(2\pi)}\tilde{g}(\omega_{1})\tilde{g}(\omega_{2})\delta(\frac{\omega_{1}}{z}+\frac{\omega_{2}}{z'})  \\
&\times J_{2\Delta+2i-1/2}(\frac{2\omega_{1}}{\lambda'})J_{2\Delta+2i-1/2}(\frac{2\omega_{2}}{\lambda'})\theta(\omega_{1}) \ .
\end{split}
\end{equation}
The sum over $i$ is the AdS generalization of the integral over $\kappa$ in \eqcref{minimizeinnerprod}. The integrals over $z$ and $z'$ correspond to a position space version of the integral over $\vec{k}$ in \eqcref{minimizeinnerprod}. Again, the integral over $\kappa$ has become a sum because putting our theory in AdS is similar to putting it in a box. The goal is to minimize again \eqcref{functionalint} for fixed $\Delta,\lambda,z$ as a functional of $g(x)$.

We restrict to the integration region $z<z'$, multiplying the final result by 2. We will extensively use the bounds \cite{NIST:DLMF}
\begin{equation}\label{boundsbessel}
\begin{split}
x>\nu:\quad J_{\nu}(x)&<\sqrt{\frac{2}{\pi \nu}}\left (\frac{x^{2}}{\nu^{2}}-1\right )^{-1/4} \\
0<x< \nu :\quad J_{\nu}(x)&<\frac{1}{\sqrt{2\pi \nu}} \left ( \frac{e x}{2\nu}\right )^{\nu} \\
\end{split}
\end{equation}
to upper bound the integral-sum in \eqcref{functionalint}. There are three integration regions to consider for $\omega_{1}$:
\begin{equation}
\begin{split}
I_{1}:& \quad \omega_{1}>\frac{\lambda'}{2}(2\Delta+2i-1/2) \\
I_{2}:& \quad \frac{\lambda'}{2}(2\Delta+2i-1/2)>\omega_{1}>\frac{z}{z'}\frac{\lambda'}{2}(2\Delta+2i-1/2) \\
I_{3}:& \quad \frac{z}{z'}\frac{\lambda'}{2}(2\Delta+2i-1/2)>\omega_{1}>0 \\
\end{split}
\end{equation}
where the upper bound we use for the Bessel functions depends on the integration domain we are in.  We denote the $I_{i}$ contribution to $B[g(x)]$ as $B_{i}[g(x)]$.

We first consider the $I_{1}$ region of integration. We replace both delta functions by the first upper bound in \eqcref{boundsbessel}, noting that we have restricted to the contour $z<z'$. The relevant integral for fixed $i$ then becomes 
\begin{equation}
\begin{split}
B_{1}[g(x)]&<\sum_{i=0}^{\infty}a_{\Delta}(i)\frac{\pi^{3}}{\lambda'}\int \frac{dz\ dz'}{(zz')^{1/2}}\int \frac{d\omega_{1}}{2\pi}\frac{\tilde{g}(\omega_{1})\tilde{g}(\frac{z'}{z}\omega_{1})}{z'^{-1}\pi\Delta_{i}}(\left ( \frac{4\omega_{1}^{2}}{\lambda'^{2}\Delta_{i}}-1\right )\left ( \frac{4z'^{2}\omega_{1}^{2}}{z^{2}\lambda'^{2}\Delta_{i}^{2}}-1\right ))^{-1/4} \\
&<\sum_{i=0}^{\infty}a_{\Delta}(i)\frac{\pi^{3}}{\lambda'}\int \frac{dz\ dz'}{(zz')^{1/2}}\int \frac{d\omega_{1}}{2\pi}\frac{\tilde{g}(\omega_{1})\tilde{g}(\frac{z'}{z}\omega_{1})}{z'^{-1}\pi\Delta_{i}}\left ( \frac{4\omega_{1}^{2}}{\lambda'^{2}\Delta_{i}^{2}}-1\right )^{-1/2} \\
\end{split}
\end{equation}
where 
\begin{equation}
\Delta_{i}=2\Delta+2i-1/2
\end{equation}
and the integration contours are $\infty>\omega_{1}>\lambda'\Delta_{i}/2$ and $0<z<z'<z_{1}+\lambda/2$. To bound this function, we use that 
\begin{equation}\label{inequalzg}
z'>z:\quad |\tilde{g}(\omega_{1})\tilde{g}(\frac{z'}{z}\omega_{1})|<\frac{z}{z'\omega_{1}^{2}}\sup_{\omega_{1}>\lambda'\Delta }|\omega_{1}g(\omega_{1})|^{2} \ .
\end{equation}
and then perform the integral over $z$ and $z'$, which simply yields a polynomial factor of $\lambda$ and $z_{1}$. We then need to bound the more non-trivial integral-sum 
\begin{equation}\label{Cfact}
c_{\textrm{AdS}}\equiv\sum_{i=0}^{\infty}\frac{a_{\Delta}(i)}{\Delta_{i}}\int \frac{d\omega_{1}}{2\pi} \frac{1}{\omega_{1}^{2}}\left ( \frac{4\omega_{1}^{2}}{\lambda'^{2}\Delta_{i}^{2}}-1\right )^{-1/2} <\infty
\end{equation}
which is the AdS${}_{2}$ generalization of \eqcref{cintflat}. To bound $c_{\textrm{AdS}}$ in \eqcref{Cfact}, we use that 
\begin{equation}
a_{\Delta'}(i)\leq \frac{\sqrt{(i+1) (2 \Delta'+i)} (2 \Delta'+2 i+1) (4 \Delta'+4 i-1)}{2 \pi  (2 i+1) (\Delta'+i)^2 (4 \Delta'+2 i-1)} \ .
\end{equation}
and that the integrand in $c_{\textrm{AdS}}$ is monotonic at large-$n$ to bound the sum by an integral. One can show it is finite and has at most polynomial behavior in $\Delta'$ at large $\Delta'$. The only non-trivial factor is $\sup_{\omega_{1}>\lambda'\Delta_{i}/2}|\omega_{1}g(\omega_{1})|^{2}$, which is upper bounded using \eqcref{upperboundgcom}. Therefore, the $B_{1}[g(x)]$ gives the sum in \eqcref{endres} which is convergent when $\lambda'>\beta_{H}$. The fact that the coordinate $\lambda'$ comes in to the result implies that there should be a more optimal derivation.

We now need to consider the $I_{2}$ and $I_{3}$ contributions which yield the ``finite'' contribution in \eqcref{endres}. Intuitively, these contributions are expected to be strongly suppressed relative to the $I_{1}$ term. At lower energies, the characteristic separation of the $\phi(x)$ insertions increases, and for large masses this increase translates into an exponential suppression as a function of the characteristic separation. In flat space, however, such suppression is insufficient to prevent infrared divergences for infinitely extended regions. In that case, 
the growth of the integration domain overwhelms the exponential decay. This pathology does not arise in AdS, where curvature effects regulate the would-be infrared divergences.
For the $I_{2}$ and $I_{3}$ contributions, we apply the upper bound on the second line of \eqcref{boundsbessel} for at least one of the Bessel functions. The second line of \eqcref{boundsbessel} trivially includes at least one factor of the form 
\begin{equation}\label{suppresionfactor}
\Delta_{i}^{-\Delta_{i}}\leq (2\Delta)^{-2\Delta} \ .
\end{equation}
Other than this factor, we find the $B_{2}[g(x)]$ and $B_{3}[g(x)]$ evaluate to functions that are rational functions of $\Delta$, $\delta$ and $z$ at large $\Delta$, so there is no infrared divergence. Therefore, the suppression factor in \eqcref{suppresionfactor} means that the $B_{2}[g(x)]$ and $B_{3}[g(x)]$ contributions are even more suppressed than $B_{1}[g(x)]$ at large $\Delta$. Importantly, even in theories with a Hagedorn spectrum, such suppression means that the $B_{2}[g(x)]$ and $B_{3}[g(x)]$ contributions do not diverge when we sum over masses. To see this, consider the sum 
\begin{equation}\label{toymodel}
\sum_{n=1}^{\infty}f(\Delta_{n})\Delta_{n}^{-\Delta_{n}}, \quad \Delta_{n}=\frac{\log(n+1)}{\beta_{H}}
\end{equation}
where $f(\Delta_{i})$ is some finite polynomial in $\Delta_{i}$. One can check via direct computation that the sum in \eqcref{toymodel} does not diverge for any choice of $f(\Delta_{i})$. 

We have so far considered only $B[g(x)]$, but the same computational strategy can, with additional effort, be applied to $\langle \Omega_{t}^{n}| j_{0}(f)^{2}|\Omega_{t}^{n}\rangle$. Introducing derivatives should not qualitatively alter the result. Dimensional analysis suggests that derivatives suppress low-energy contributions to the integral while enhancing high-energy ones. Since we have already shown that the integral is infrared finite even without derivatives, there is no genuine concern at low energies. The main challenge is to ensure that the enhanced high-energy behavior remains under control. In flat space, this was indeed the case. Because the ultraviolet regime of the integral is expected to be insensitive to the large-scale structure of spacetime, the flat-space analysis strongly suggests that derivatives should likewise be unimportant in AdS. More quantitatively, we can use the identity
\begin{equation}\label{signide}
4|\partial_{t_{1}}^{2}\langle \phi(z,t_{1})\phi(z',t_{2})\rangle^{2}| \geq |[(\partial_{t_{1}}-\partial_{t_{1}'})^{2}\langle \phi(z,t_{1})\phi(z',t_{2})\rangle\langle \phi(z',t_{1}')\phi(z'',t_{2}')\rangle]|_{t_{i}'\rightarrow t_{i},z_{i}'\rightarrow z_{i}}|
\end{equation}
to eliminate the time derivatives. To see that \eqcref{signide} holds, one should convert each propagator in \eqcref{signide} to a momentum space representation in \eqcref{momentumspaceAdS}. \eqcref{signide} then follows from the identity
\begin{equation}
\omega,\omega'\geq 0: \quad (\omega+\omega')^{2}\geq (\omega-\omega')^{2}
\end{equation}
where $\omega$ ($\omega'$) is the conjugate variable to $t_{1}$ ($t_{1}'$) of the first (second) propagator in the product. The bounds on $\omega$ and $\omega'$ follow from the Heaviside function in the momentum representation of the bulk propagator. We then repeat the previous proof, except we have additional factors of $\omega$ due to the time derivatives in \eqcref{signide}. These additional factors of $\omega$ do not qualitatively change the underlying argument and do not quantitatively change the final result.

\bibliographystyle{JHEP}
    \bibliography{main}
\end{document}